\documentclass[preprint,12pt]{elsarticle}

\usepackage{amssymb,amsmath,mathtools}
\usepackage{amsthm}
\usepackage{graphicx,color,subfigure}
\usepackage{xcolor}
\usepackage{algorithm}
\usepackage{algpseudocode}
\usepackage{tikz}
\usetikzlibrary{arrows.meta,positioning}
\usepackage{multirow}
\usepackage{diagbox}
\usepackage{nomencl,framed}
\usepackage{multicol}
\usepackage[capitalize]{cleveref}
\usepackage{dsfont}
\usepackage[printonlyused]{acronym}
\biboptions{sort&compress}

\newtheorem{remark}{Remark}

\makenomenclature

\newtheorem{theorem}{Theorem}
\newtheorem{lemma}{Lemma}

\newcommand{\rangeT}{\tau}
\newcommand{\I}{\mathrm{I}}
\newcommand{\X}{\mathcal{X}}
\newcommand{\Yu}{\mathcal{X}^\bu}
\newcommand{\Yp}{\mathcal{X}^\bp}
\newcommand{\Yui}[1]{\mathcal{X}^{\bu,#1}}
\newcommand{\Ypi}[1]{\mathcal{X}^{\bp,#1}}
\newcommand{\mX}{\mathrm{X}}
\newcommand{\T}{\mathcal{T}}
\newcommand{\Ti}[1]{\mathcal{T}^{#1}}
\newcommand{\N}{\mathbf{N}}
\newcommand{\G}{\mathrm{G}}
\newcommand{\D}{\mathrm{D}}
\newcommand{\mT}{\mathrm{\bf T}}
\newcommand{\bT}{\mathrm{ T}}
\newcommand{\A}{\mathrm{A}}
\newcommand{\V}{\mathrm{V}}
\newcommand{\Q}{\mathrm{Q}}
\newcommand{\M}{\mathrm{M}}
\newcommand{\mM}{\mathrm{M}^h}
\newcommand{\mK}{\mathrm{K}^h}
\newcommand{\bu}{\mathbf{u}}
\newcommand{\bp}{\mathbf{p}}
\newcommand{\Xu}{\mathrm{X}^\bu}
\newcommand{\Xp}{\mathrm{X}^\bp}
\newcommand{\bx}{\mathbf{x}}

\newcommand{\bus}{\bu^\star}
\newcommand{\bxs}{\bx^\star}
\newcommand{\bps}{\bp^\star}
\newcommand{\ts}{t^\star}

\newcommand{\tu}{\mathbf{u}^{\rm TSE}}
\newcommand{\tp}{\mathbf{p}^{\rm TSE}}
\newcommand{\pu}{\mathbf{u}^{\rm PGD}}

\newcommand{\spu}{\mathbf{u}^{\rm SPGD}}
\newcommand{\spp}{\mathbf{p}^{\rm SPGD}}

\newcommand{\uk}{\mathrm{u}}
\newcommand{\huk}{\hat{\mathrm{u}}}
\newcommand{\pk}{\mathrm{p}}
\newcommand{\ukh}{\mathrm{u}^h}
\newcommand{\rankm}{\mathrm{m}}
\newcommand{\n}{n}
\newcommand{\nn}{N}
\renewcommand{\d}{\mathrm{d}}
\newcommand{\OPX}[2]{\left[\int_{\Omega}#1\cdot#2 \,\d \bx\right]}
\newcommand{\mpk}[2]{\mathcal{M}_{#1}^{#2}}
\newcommand{\cpk}[2]{\mathcal{C}_{#1}^{#2}}
\newcommand{\qpk}[2]{\mathcal{Q}_{#1}^{#2}}
\newcommand{\npk}[2]{\mathcal{N}_{#1}^{#2}}
\newcommand{\ppk}[2]{\mathcal{P}_{#1}^{#2}}
\newcommand{\epk}[2]{\epsilon_{#1}^{#2}}
\newcommand{\gpk}[2]{\gamma_{#1}^{#2}}
\newcommand{\vepk}[2]{\xi_{#1}^{#2}}
\newcommand{\vpk}[2]{\varepsilon_{#1}^{#2}}
\newcommand{\dpk}[2]{\delta_{#1}^{#2}}
\newcommand{\apk}[2]{a_{#1}^{#2}}
\newcommand{\stab}{\beta}
\newcommand{\staba}{\alpha}
\newcommand{\bnp}[2]{b_{#1}^{#2}}
\newcommand{\wnp}[2]{w_{#1}^{\,#2}}
\newcommand{\vnp}[2]{y_{#1}^{#2}}
\newcommand{\Anp}[2]{A_{#1}^{#2}}
\newcommand{\Bnp}[2]{B_{#1}^{#2}}
\newcommand{\nod}[2]{\mathrm{n}_{#1}^{#2}}

\renewcommand{\P}{\mathcal{P}}

\newcommand{\ftse}{f_{\rm TSE}}

\newcommand{\Ftse}{F_{\rm TSE}}
\newcommand{\Gtse}{G_{\rm TSE}}
\newcommand{\Fpgd}{F_{\rm PGD}}
\newcommand{\Gpgd}{G_{\rm PGD}}
\newcommand{\Falt}{F_{\rm SPGD}}
\newcommand{\Galt}{G_{\rm SPGD}}
\newcommand{\Fns}{\mathrm{H}}
\newcommand{\Fif}{\mathrm{I}}
\newcommand{\Gif}{J}
\newcommand{\Gns}{\mathrm{J}}
\newcommand{\bn}{\mathbf{n}}

\newcommand{\Rey}{\mathrm{Re}}

\DeclareMathOperator*{\argmin}{arg\,min}

\begin{document}

\begin{frontmatter}

\title{From Time Series Expansion to Proper Generalized Decomposition via Graph-Theoretical Connection: Stabilized Simulation of Fluids Flow}

\author[label1]{Ahmad Deeb\corref{cor1}}
\ead{ahmad.deeb@ku.ac.ae}
\author[label2]{Vladimir Parezanović}
\ead{vladimir.parezanovic@ku.ac.ae}
\author[label1]{Denys Dutykh}
\ead{denys.dutykh@ku.ac.ae}

\affiliation[label1]{organization={Khalifa University of Science and Technology, Mathematics Department}, addressline={PO Box 127788}, city={Abu Dhabi},
            country={United Arab Emirates}}

\affiliation[label2]{organization={Khalifa University of Science and Technology, Aerospace Engineering Department}, addressline={PO Box 127788}, city={Abu Dhabi},
            country={United Arab Emirates}}

\cortext[cor1]{Corresponding author}

\begin{abstract}
In this paper, we employ graph theory to establish a connection between the \ac{TSE} and \ac{PGD} methods. Using the concept of a directed graph, we demonstrate how one can transition from the computation of space modes in the \ac{TSE}—first illustrated for the diffusion equation—to those of space modes in \ac{PGD}, in which an inhomogeneous Volterra-type convolution recurrence relation, weighted by time-dependent coefficients, appears.
This recurrence relation is simplified through graph-based analysis into a compact form using a simple path traversal, reducing the computational complexity. Moreover, the compact formulation reveals a natural stabilization process in the computation of space modes, where stabilized coefficients are automatically derived and can be used in the \ac{STSE} framework.
To explicitly construct these coefficients, we consider a \ac{SPGD} formulation in which the time modes are chosen as a polynomial basis $t^n$. This choice yields a one-level Volterra-type recurrence relation that is similarly simplified using a simple path representation, demonstrating a connection in the computation of space modes from \ac{TSE}, through \ac{STSE} and \ac{SPGD}, to \ac{PGD}.
This graph-based connection is exhibited in the case of inviscid flow to check how crucial the addition of an artificial diffusion is in stabilizing the recurrence formula of \ac{TSE}.
Finally, we extend the approach to the incompressible, dimensionless \ac{NS} equations and build stabilization coefficients that depend on the Reynolds number $\Rey$, the space mode rank, and the simulation time step. Both the \ac{STSE} and \ac{SPGD} approaches are tested to simulate the wake behind a bluff body at $\Rey = 5\,000$.
\end{abstract}

\begin{highlights}
	\item A novel graph-theoretic framework is introduced to connect \ac{TSE} and \ac{PGD} methods.
	\item A compact recurrence formula simplifies \ac{PGD}'s space mode computation via direct paths.
	\item Close forms of stabilization coefficients are provided for \ac{STSE} method.
	\item The proposed method is applied to simulate bluff body wake at $\Rey = 5000$.
\end{highlights}


\end{frontmatter}

\section{Introduction}
The numerical solution of unsteady incompressible \acf{NS} equations remains a significant challenge in both computational fluid dynamics and numerical analysis. With the rise of Machine Learning for solving \ac{PDE} \cite{ML_PDE, ZHANG2024116936, SOUSA2024117133, CARON2025112229}, numerical simulations based approximations remain a key to understanding natural phenomena \cite{QI_turb, MAO2024117172, VALIZADEH2025117618, SHI2026118412, RATH2024117348, LARA2025117745}.

Time integration schemes \cite{IBRAHIMBEGOVIC20024241, PARK2011130,   deeb:casson,deeb:AML1}, especially when combined with high-order \acf{FEM} \cite{SOLIN20121635, GARCIAESPINOSA2015290}, often suffer from instabilities and stiffness, particularly when high accuracy \cite{WESTERMANN2024116545} or long-time integration \cite{ahmad_robust_integrators_2019} are required.
In a recent study \cite{Deeb-NS-25}, Deeb et \emph{al.} have developed a stabilized FEM framework that is tailored for a time integration scheme of \ac{NS} equations that is based on the \acf{DSR} \cite{ahmad_bpl_2014,ahmad_comp_bpl_sfg_2015, ahmad_icnpaa_2016, DEEB_2022_bpl} technique. Within this framework, the solution is expressed in the form of \acf{TSE}, and upon substitution into the governing equations, a cascade of coupled equations for the spatial modes of velocity and pressure is obtained. These equations are solved recursively, each spatial mode depending on the previous ones.

However, the computation of high-rank spatial modes using high-order \ac{FEM} reveals numerical instabilities. In contrast with stabilization methods \cite{Berger2015A2222, Tezduyar_1991} such as the \ac{SUPG} method applied for convection dominated flows \cite{brooks_supg_1982} and the \ac{PSPG} \cite{ARONSON2025117633} circumventing the Babuška-Brezzi condition \cite{hughes_BB_1986}, a stabilization technique was proposed first in \cite{deeb:stab-serie} to mitigate this by introducing an artificial diffusion term in the left-hand side of the recursive equations. The stabilization coefficient associated with the first rank spatial mode is computed offline, prior to the time integration, by minimizing the condition number of the resulting mass matrix. This optimization was shown to be aligned with the \ac{DMP} in the case of the transient Laplace equation and allowed for significantly larger time steps, up to two orders of magnitude larger, compared to the unstabilized case.
However, this process remains costly and heuristic, especially when tuning coefficients for higher-rank modes, which still relies on feedback from numerical experiments.

On the other hand, the \ac{PGD} method \cite{ladeveze_1999,chinesta_2011} is well-known for its robustness in solving parametric and high-dimensional problems, offering separated representations in space-time \cite{Ammar_2007_1, Ammar_2007_2, FERNANDES2021114102} and parameters \cite{CHINESTA2011578,ahmad_pgd_pade}. In this work, we aim to explore a fundamental connection between the \ac{TSE} and \ac{PGD} frameworks through the lens of graph theory \cite{graph_theory_2025}. Our goal is to develop a more efficient and analytically grounded method for determining the coefficient of stabilization of \ac{TSE} that avoids the costly offline minimization and can adapt naturally to the structure of the problem.

Beyond the development of stabilized and graph-based methods, this work also addresses the practical application of these approaches to simulate the unsteady wake behind a two-dimensional bluff body. The accurate numerical prediction of bluff body wakes is essential for many engineering applications and has been the subject of extensive investigation. Experimental studies such as \cite{PRASAD_3dwakes_1997} revealed the complexity of bluff body wake dynamics, including three-dimensional instabilities that influence the evolution of higher-order flow structures, which are often difficult to capture in two-dimensional simulations. \ac{DNS} at Reynolds number $\Rey = 5000$, such as those conducted in \cite{rodriguez_DNS_2011}, provide a high-fidelity benchmark with detailed turbulence statistics, serving as a valuable reference for assessing numerical methods. Additionally, the combined use of \ac{LES} and \ac{PIV} in \cite{yagmur_2014} offers both numerical and experimental insights into the behaviour of wakes at this Reynolds number. These studies highlight the importance and difficulty of accurately resolving the dominant flow features in the wake, particularly in terms of drag, lift, and the coherence of POD modes. The present work contributes to this body of literature by applying and evaluating the \ac{STSE} and \ac{SPGD} methods in the simulation of the bluff body wake at $\Rey = 5000$.

The paper is structured as follows: The concept of summing the recurrence formula in graph theory is presented in \cref{sec2}. Then, the connection between \ac{TSE} to \ac{PGD}, passing by \ac{STSE} and \ac{SPGD}, is presented in \cref{sec3}. In \cref{sec40}, we investigate the connection for inviscid flow, to see the effect of adding artificial diffusion in the left-hand side of the recurrence formula. In \cref{sec4}, we will extend the connection to the \ac{NS} equations and apply these methods to simulating the wake of a two-dimensional bluff body in \cref{sec5}. We end this paper with conclusions in \cref{sec6}.

\section{Graph theory background}
\label{sec2}
Let us consider the following recursive formula:
\begin{equation}
 \label{recursive_fN}
\begin{aligned}
 f(\n) &= \ g(\n) + \sum\limits_{p=1}^{\n-1} f(p)\,\wnp{p}{\n}, &  \forall \n\geqslant 1.
 \end{aligned}
\end{equation}
This recurrence formula is a Volterra-type convolution over the function $f(p)$ weighed by the coefficients $\wnp{p}{\n}$. The second kind Volterra integral equation is a discrete linear form for having a known function $g(\n)$ representing the inhomogeneous part of the recurrence formula.

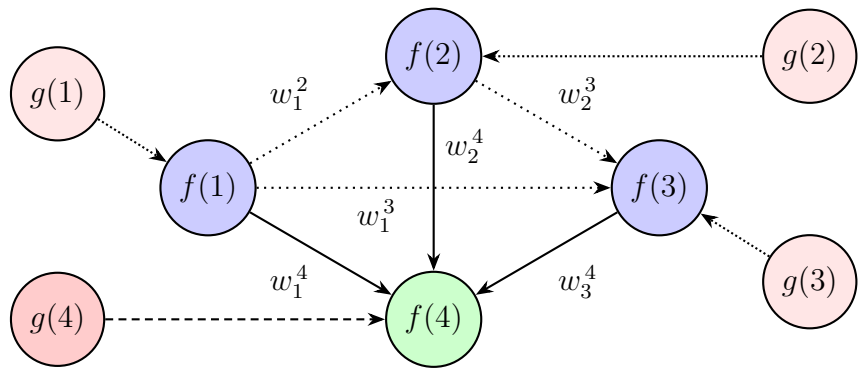
\begin{figure}[h!]
 \centering
\begin{tikzpicture}[>=Stealth, node distance=2.5cm, thick]

  \node[circle, draw, fill=red!20] (g4) at (-2,-4) {$g(4)$};
  \node[circle, draw, fill=blue!20] (n1) at (0,-2.25) {$f(1)$};
  \node[circle, draw, fill=blue!20] (n2) at (3,-0.5) {$f(2)$};
  \node[circle, draw, fill=blue!20] (n3) at (6,-2.25) {$f(3)$};
  \node[circle, draw, fill=green!20] (n4) at (3,-4) {$f(4)$};

  \draw[->] (n1) -- (n4) node[midway, below left] {$\wnp{1}{4}$};
  \draw[->] (n2) -- (n4) node[near start, right] {$\wnp{2}{4}$};
  \draw[->] (n3) -- (n4) node[midway, below right] {$\wnp{3}{4}$};

  \draw[->, densely dashed] (g4) -- (n4) node[midway, above left] {};

  \draw[->, dotted] (n1) -- (n2) node[midway, above left] {$\wnp{1}{2}$};
  \draw[->, dotted] (n1) -- (n3) node[ near start, below right] {$\wnp{1}{3}$};
  \draw[->, dotted ] (n2) -- (n3) node[midway, above right] {$\wnp{2}{3}$};

  \node[circle, draw, fill=red!10] (g1) at (-2,-1) {$g(1)$};
  \node[circle, draw, fill=red!10] (g2) at (8,-0.5) {$g(2)$};
  \node[circle, draw, fill=red!10] (g3) at (8,-3.5) {$g(3)$};
  \draw[->, densely dotted] (g1) -- (n1) node[midway, above left] {};
  \draw[->, densely dotted] (g2) -- (n2) node[midway, above left] {};
  \draw[->, densely dotted] (g3) -- (n3) node[midway, above left] {};

\end{tikzpicture}
\caption{Graph visualization of the recurrence formula in \cref{recursive_fN} for $\n=4$. \label{fig1_graph5}}
\end{figure}

Formula \eqref{recursive_fN} is a one-level memory $\n$; \emph{i.e.} depending on one parameter $\n$. It can be visualized using a direct graph, where each value $f(p),\, p=1,\ldots \n,$ represents the value at the node $p$ in this graph while direct edges from node $p$ to $\n$ represent the contribution of node $p$ to node $\n$ weighted by $\wnp{p}{\n}$.  \cref{fig1_graph5} represents the graph for $\n=4$ where solid lines represent the direct contribution to node $\n$. Note that $f(2)$ and $f(3)$ are already obtained, and dotted lines represent their graph. We write the recurrence formula of $f(1)$ and $f(2)$:
\begin{equation*}
 \label{recursive_f21}
 \begin{aligned}
  f(1) &= g(1), \\
  f(2) &= g(2) +  f(1)\wnp{1}{2} = g(2) + g(1) \wnp{1}{2},\\
 \end{aligned}
\end{equation*}
which are used to express $f(3)$ as follows:
\begin{equation*}
 \label{recursive_f2}
 \begin{aligned}
  f(3) & = g(3) + f(1)\wnp{1}{3}+ f(2)\wnp{2}{3},\\
   &= g(3)  +g(1)\wnp{1}{3} + \big(g(2) + g(1) \wnp{1}{2} \big)\wnp{2}{3},\\
   & = g(3) + g(2)\wnp{2}{3} + g(1)\big(\wnp{1}{3} +\wnp{1}{2}\wnp{2}{3}  \big).
 \end{aligned}
\end{equation*}
In the last line, we assembled all the weights multiplying $g(p)$ and ordered them, after replacing $f(2)$ by its formula. We replace now all $f(p), p=1,2,3$ in the formula of $f(4)$, and assemble all the terms of $g(p),p=1,\ldots,4$ to obtain:
\begin{equation*}
\begin{aligned}
 f(4) &= g(4) + f(3)\wnp{3}{4} + f(2)\wnp{2}{4} + f(1)\wnp{1}{4},\\
 &= g(4) + g(3)\wnp{3}{4} + g(2)\big(\wnp{2}{4} + \wnp{2}{3}\wnp{3}{4})\\
 & + g(1)\big(\wnp{1}{4} + \wnp{1}{2}\wnp{2}{4} + \wnp{1}{3}\wnp{3}{4} + \wnp{1}{2}\wnp{2}{3}\wnp{3}{4}).
\end{aligned}
\end{equation*}

Using the concept of direct graph, we can formulate $f(\n)$ using a simple path from $1$ to $\n$ as follows:
\begin{equation}
\label{rec_fromula}
 f(\n) = g(\n) + \sum\limits_{p=1}^{\n-1} g(p) \left( \sum\limits_{P \in \P(p,\n)} \,\,\prod\limits_{(i,j) \in P} \wnp{i}{j} \right),
\end{equation}
where $\P(p,\n)$ is the set of all simple path from node $p$ to node $\n$, $P$ is a path in this set, $(i,j)$ is an edge in this path. The last formula can be written in a compact form:
\begin{equation}
 \label{formula_fn}
 f(\n) = \sum\limits_{p=1}^{\n} \psi(p,\n)\, g(p),
\end{equation}
with $\psi(\n,\n)=1$ and verifying the recurrence formula:
\begin{equation}
\label{formula_psi}
\psi(p,n) = \sum\limits_{q=p}^{\n-1} \psi(p,q)\, \wnp{q}{\n}= \sum\limits_{P \in \P(p,\n)} \,\,\prod\limits_{(i,j) \in P} \wnp{i}{j},\quad \forall p<\n.
\end{equation}
We present the pseudo-algorithm computing $\bnp{p}{\n}\coloneqq \psi(p,n)$ in \cref{Alg_bnp} presented in \cref{append_alg_bnp}. This algorithm builds a Direct Graph with nodes, edges, and weights, computes the sum of weights for every path from $p$ to $\n$, and stores the sum in $\bnp{p}{\n}$. In \textsc{Python} language, we are using the package \textsc{Networkx} and the function \textsc{Networkx.DiGraph} to create a direct graph $DiGr$. Adding edges for node $p$ to $\n$ weighed by $\wnp{p}{\n}$ is done with the sub-function \textsc{G.add\_edges($p,\n,\wnp{p}{\n}$)}.

In addition to the importance of Formula \eqref{rec_fromula} in studying the asymptotic behaviour of the series governed by, this compact representation of $f(\n)$ is important as it gives a straightforward way to compute the value of $f(\n)$ by only using $g(\n)$, which will be shown later in the numerical simulation.

\section{Connection between \ac{PGD} and \ac{TSE} for the diffusion problem}
\label{sec3}
To simplify the establishment of the connection for the reader, we will consider first the following linear diffusion equation:
\begin{equation}
 \label{diff-eq}
 \partial_t \bu = \nu \Delta \bu, \quad \bx \in \Omega \subset \mathds{R}^d,
\end{equation}
where $t$ is the time scalar variable, $\partial_t$ is the partial derivative relative to time, $\nu$ is the diffusion coefficient, and $\Delta$ is the Laplace operator relative to the vector of space variables $\bx$. Having fixed $\nu$, the \ac{PGD} consists of writing the solution as a sum of tensor products as follows:
\begin{equation}
\label{u_pgd}
\pu( t, \bx) \coloneqq  \sum\limits_{\n=0}^{\infty}\left(\T_\n  \otimes\X_n  \right) ( t, \bx) = \sum\limits_{\n=0}^{\infty} \T_\n(t)\cdot \X_\n(\bx),
\end{equation}
where functions $\T_\n(t)$, considered to be in $ L^2([0,\rangeT])$, and $\X_\n(\bx) \in \big(H^1(\Omega)\big)^d$ are both found in a recursive way, being orthogonal to the residual of the \cref{diff-eq}. One has $\X_p$ and $\T_p$ up to rank $p=\n-1$, elements $\T_\n$ and $\X_\n$ are to be found by replacing the following finite sum
\begin{equation}
 \pu_\n(t,\bx) \coloneqq \sum\limits_{p=0}^{\n}\T_p(t)\cdot \X_p(\bx),
\end{equation}
as approximation of $\bu$ in \cref{diff-eq} to obtain the equation:
\begin{equation}
\label{diff-pgd}
 \sum\limits_{p=0}^{\n}{\T_p}^{\prime}(t)\cdot \X_p(\bx) = \nu\sum\limits_{p=0}^{\n}\T_p(t)\cdot \Delta \X_p(\bx),
\end{equation}
where $\T_p^{\prime}$ is the time derivative of the time mode $\T_p(t)$.
On the other hand, the solution $\bu(t,x)$ can be written in its \ac{TSE} as follows:
\begin{equation}
\label{u_tse}
 \tu(t,x) \coloneqq \sum\limits_{k=0}^{\infty} \uk_k(\bx) \, t^k.
\end{equation}
After injecting it in \cref{diff-eq} and equating terms of $t^k$, we obtain the recurrence formula of the space modes $\uk_k(x)$ as follows:
\begin{equation}
 \label{uk}
 \uk_{k+1}(x) = \frac{\nu}{k+1}\Delta \uk_{k}(x), \quad \forall k>0, \quad \text{and} \quad \uk_0(\bx) = \bu(0,\bx).
\end{equation}

\subsection{Finding {$\uk_k$}}
\begin{figure}[htp]
 \centering
\begin{tikzpicture}[>=Stealth, node distance=1.9cm, every node/.style={scale=0.9}]
  \node[circle, draw, fill=blue!10] (f0) {$f(0)$};
  \node[circle, draw, fill=blue!10] (f1) [right=of f0] {$\ftse(1)$};
  \node[circle, draw, fill=blue!10] (f2) [right=of f1] {$\ftse(2)$};
  \node[circle, draw, fill=blue!10] (f3) [right=of f2] {$\ftse(3)$};

  \draw[->,dotted] (f0) -- (f1) node[midway, above] {$\frac{\nu}{1}\Delta$};
  \draw[->,dotted] (f1) -- (f2) node[midway, above] {$\frac{\nu}{2}\Delta$};
  \draw[->] (f2) -- (f3) node[midway, above] {$\frac{\nu}{3}\Delta$};

\end{tikzpicture}
\caption{Graph representation of recurrence formula $\uk_3$. \label{fig1_graphuk}}
\end{figure}
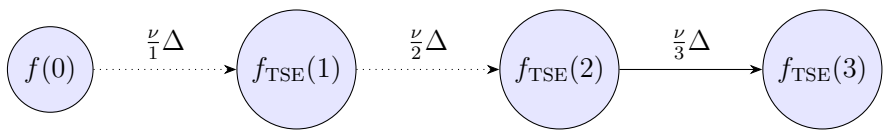

The recurrence formula \eqref{uk} can be written in a different way, where $\uk_k$ is the solution of:
\begin{equation}
 \label{Ftse0}
 \Ftse(k+1)\coloneqq (k+1)\uk_{k+1} - \nu \Delta \uk_{k} = 0,
\end{equation}
while the series terms $\ftse(k)(x) \coloneqq \uk_k(\bx)$ can be represented by the graph in \cref{fig1_graphuk} and verifying a functional recurrence:
\begin{equation}
\label{uk_fn}
\ftse(k)=  g_k\circ \ftse(k-1), \quad \text{where} \quad g_k = \frac{\nu}{k} \Delta.
\end{equation}

If $\uk_0$ is known in its closed form, we can build a sequence of functions $\ftse(k)$ where each function is the Laplacian of the previous one, scaled by $\frac{\nu}{k}$. The recurrence formula could be written as a function of $\ftse(0) =  \uk_0$ as follows:
\begin{equation}
 \ftse(k)(\bx)  = \big( g_k\circ\ldots \circ g_1\big)\circ\ftse(0)= \frac{\nu^k}{k!} \Delta^ k \ftse(0),
\end{equation}
for which the sum of $\ftse(k)$ represents the Taylor expansion of the Gaussian kernel acting on $\uk_0$.
If one has found the terms up to a finite rank $\n$, the approximation is close to the exact solution with a residual error of:
\begin{equation}
 \sum\limits_{k=\n+1}^{\infty} \ftse(k) \, t^k.
\end{equation}
This recurrence formula is impractical numerically as discussed in \cite{deeb:stab-serie}. Thus, numerical computation using numerical methods such as the \acf{FDM} or \acf{FEM} is to be employed. However, it was also shown that the computed spatial modes via \cref{uk} in the framework of high-order \ac{FE} are not reliable and explode when the condition number of the Mass matrix does not fulfil the \ac{DMP}. Mass lumping improves the stability for the first order \ac{FE} but only reduces the errors and does not affect its asymptotic behaviour when applied to the second and higher orders. Thus, a stabilization technique was proposed in \cite{deeb:stab-serie}. Instead, $\uk_{k}$ has to verify an updated recurrence equation:
\begin{equation}
\begin{aligned}
 \label{Ftse}
 \Ftse(k) &=\Gtse(k)& \text{with} \quad
 \Gtse(k) \coloneqq \staba_{k-1} \Delta \uk_k,\,\,\, k\geqslant 1,
 \end{aligned}
\end{equation}
and the recurrence formula generating $\uk_k(x) = \ftse(k)$ is given:
\begin{equation}
\label{stab_uk}
 \ftse(k)
 = \tilde g_k \circ \ftse(k-1), \,\,\, \text{where}\,\,\, \tilde g_k = \frac{\nu}{k} \left[\I - \staba_{k-1} \Delta \right]^{-1} \circ \Delta,\,\,\, k\geqslant 1,
\end{equation}
which is referred to as the \ac{STSE} method. After assembling the equation within \ac{FEM}, we present its discrete form below:
\begin{equation}
 \label{discrete_stab_uk}
  \ukh_{k} = -\frac{\nu}{k} \Big( \mM+\staba_{k-1} \mK \Big)^{-1} \cdot\mK \ukh_{k-1}.
\end{equation}
The coefficient $\staba_0$ is sought to be the $\argmin$ of the condition number $\kappa(\mM+\staba \mK)$ for which it was proven in \cite{deeb:stab-serie} that the maximum principle is preserved.
Numerical experiments showed that
\begin{equation}
\staba_0 \sim O(h^{m}),
\end{equation}
where $h$ is the size of the mesh and $m \sim 2 + o(h,s)$ is a coefficient that slightly change with the mesh size $h$ and  the \ac{FE} order $s$: it decreases when $s$ increases or $h$ decreases. For $k\geqslant 1$, $\staba_k$ might follow empirical formula such as:
\begin{align}
 \staba_{k+1} &= \staba_0, & \staba_{k+1} &= \mathrm{C}\cdot \staba_k,& \forall k\geqslant 0,
\end{align}
where $\mathrm{C}$ is an empirical constant that is usually taken as $2$ also. This choice is defended by the fact that the oscillations are amplified in the mesh at every rank $k$. One of the advantages of building the connection with \ac{PGD} is to provide a close form of $\staba_k$.

\begin{remark}
The operator $\left[\I - \staba \Delta \right]^{-1}$ is a Helmholtz filter operator, and acts as a low-pass filter by damping small-scale variations while preserving large-scale structures. This filter appears when the stabilization technique is applied to compute the spatial modes of the \ac{TSE} as shown above. The size of the filter depends on the coefficient $\staba$.
\end{remark}

\subsection{Finding $\T_\n$}
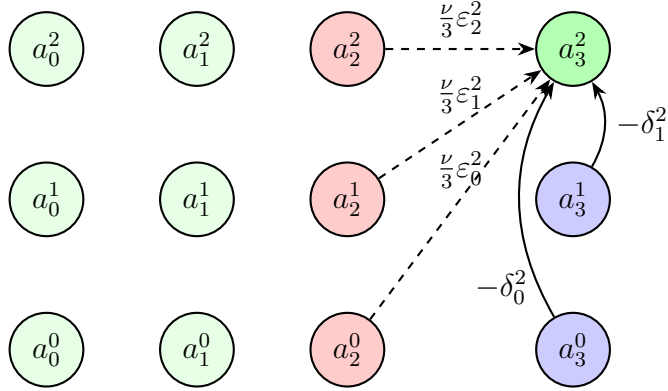
\begin{figure}[h!]
 \centering
\begin{tikzpicture}[>=Stealth, thick, node distance=2cm]
  \node[circle, draw, fill=green!10] (f00) at (0, 0) {$\apk{0}{0}$};
  \node[circle, draw, fill=green!10] (f10) at (2, 0) {$\apk{1}{0}$};
  \node[circle, draw, fill=red!20] (f20) at (4, 0) {$\apk{2}{0}$};
  \node[circle, draw, fill=blue!20]   (f30) at (7, 0) {$\apk{3}{0}$};

  \node[circle, draw, fill=green!10] (f01) at (0, 2) {$\apk{0}{1}$};
  \node[circle, draw, fill=green!10] (f11) at (2, 2) {$\apk{1}{1}$};
  \node[circle, draw, fill=red!20] (f21) at (4, 2) {$\apk{2}{1}$};
  \node[circle, draw, fill=blue!20] (f31) at (7, 2) {$\apk{3}{1}$};

  \node[circle, draw, fill=green!10] (f02) at (0, 4) {$\apk{0}{2}$};
  \node[circle, draw, fill=green!10] (f12) at (2, 4) {$\apk{1}{2}$};
  \node[circle, draw, fill=red!20] (f22) at (4, 4) {$\apk{2}{2}$};
  \node[circle, draw, fill=green!30]   (f32) at (7, 4) {$\apk{3}{2}$};

  \draw[->,dashed] (f20) -- (f32) node[midway, above] {$\frac{\nu}{3} \vpk{0}{2}$};
  \draw[->,dashed] (f21) -- (f32) node[midway, above] {$\frac{\nu}{3} \vpk{1}{2}$};
  \draw[->,dashed] (f22) -- (f32) node[midway, above ] {$\frac{\nu}{3} \vpk{2}{2}$};
  \draw[->,] (f30) to[bend left=30] node[very near start, left] {$- \dpk{0}{2}$} (f32);
  \draw[->,] (f31) to[bend right=30] node[midway, right] {$- \dpk{1}{2}$} (f32);
\end{tikzpicture}
 \caption{Figure of the graph generating the coefficients $\apk{k}{\n}$ .\label{fig1_graphnk}}
\end{figure}

If we consider that $\X_n$ is already known, thus the function $\T_n$ is found as follow: We multiply $\X_n$ with the \cref{diff-pgd} and integrate over the domain $\Omega$ to obtain the \ac{ODE} governing $\T_\n$:
\begin{equation}
\label{ODE_Tn}
\mpk{\n}{\n}\cdot\T_\n^{\prime}(t)  =  \sum\limits_{p=0}^{\n-1} \nu \qpk{p}{\n} \cdot \T_p(t)-\mpk{p}{\n}\cdot\T_p^{\prime}(t) + \nu \qpk{\n}{\n} \cdot \T_\n(t),
\end{equation}
where the coefficients $\mpk{p}{\n}$ and $\qpk{p}{\n}\in \mathds{R}$ and are given below:
\begin{align}
 \mpk{p}{\n} & = \OPX{\X_p(\bx)}{\X_\n(\bx)}, &\qpk{p}{\n} & = \OPX{\Delta \X_p(\bx)}{\X_\n(\bx)}, \quad \forall p \leqslant \n.
\end{align}
To complete the \ac{IVP} of \cref{ODE_Tn}, the initial condition $\T_\n(0)\in \mathds{R}$ is generated using the initial condition of \cref{diff-eq} as follow:
\begin{equation}
 \label{T_n0}
 \T_\n(0) = \frac{1}{\mpk{\n}{\n}} \left[\OPX{\bu(0,\bx)}{\X_\n(\bx)} - \sum\limits_{p=0}^{\n-1} \T_p(0) \mpk{p}{\n} \right].
\end{equation}
The unknown $\T_p(t)$ could be written in its \ac{TSE} form as follows:
\begin{equation}
\label{Tp_tse}
 \T_p(t) = \sum\limits_{k=0}^{\infty} \apk{k}{p}\, t^k ,\quad \apk{k}{p}\in \mathds{R}, \quad \forall p \geqslant 0,
\end{equation}
with $\apk{0}{p} = \T_p(0)$ are the initial conditions of \ac{IVP} governing $\T_p(t)$. After replacing $\T_p$ and their derivatives for $p=0,\ldots,\n$ in \cref{ODE_Tn}, we obtain the recurrence formula generating coefficients $\apk{k}{\n}$ as follows:
\begin{equation}
\label{apkpgd}
 \apk{k}{\n}= \frac{\nu}{k}\sum\limits_{p=0}^{\n}  \vpk{p}{\n} \apk{k-1}{p} - \sum\limits_{p=0}^{\n-1} \dpk{p}{\n} \apk{k}{p},
\quad \forall k\geqslant 1, \forall \n\geqslant 1,
\end{equation}
where coefficients $\vpk{p}{\n}$ and $\dpk{p}{\n}$ are given below:
\begin{align}
\vpk{p}{\n} &= \frac{\qpk{p}{\n}}{\mpk{\n}{\n}}, &
\dpk{p}{\n} &= \frac{\mpk{p}{\n}}{\mpk{\n}{\n}}.
\end{align}
The recursive formula in \cref{apkpgd} is a two-level memory (horizontal in $k$ and vertical in $\n$), which is non-homogeneous in the $k$ level.
The first sum involves the nodes related to the last level $k-1$ (memory in $k$) scaling by a factor $k$, and the second sum involves all the last nodes from $p=0$ to $\n-1$ (memory in $\n$) in the level $k$. This is illustrated in \cref{fig1_graphnk} when computing $\apk{3}{2}$. Note that $\apk{k}{p},\, p=1,\ldots \n-1$, for any $k$, are mandatory for $\apk{k}{\n}$ as $\T_\n(t)$ is found consecutively from $ p=1,\ldots \n-1$ and in a recursive way with $\X_\n(x)$.
If we denote by $f(k,\n)\coloneqq \apk{k}{\n}$, \cref{apkpgd} can be written as follows:
\begin{equation}
 \label{apkpgdfnk}
 \begin{aligned}
 f(k,\n) &=  g_k\left(\sum\limits_{p=0}^{\n}  f(k-1,p) \Anp{p}{\n} \right)+ \sum\limits_{p=0}^{\n-1} f(k,p) \Bnp{p}{\n},\quad g_k(f) =\frac{\nu}{k}\cdot f,
\end{aligned}
 \end{equation}
 with $\Anp{p}{\n} = \vpk{p}{\n}$ and $\Bnp{p}{\n} = - \dpk{p}{\n}$.

The illustration of the recurrence formula \eqref{apkpgdfnk} of two levels, $\n$ and $k$, in the graph in \cref{fig1_graphnk} is helpful to express $\apk{k}{\n}$ in function of $\apk{r}{0},\, r=0,\ldots,k,$ and $\apk{0}{p},\,p=1,\ldots,\n,$  as follows:
\begin{equation}
 \label{apkpgdf0}
 \apk{k}{\n} = \frac{\nu^k}{k!} \left[\sum\limits_{p=1}^{\n} \phi\big(\nod{0}{p},\nod{k}{\n}\big)\,\apk{0}{p} + \sum\limits_{r=0}^{k} \frac{r!}{\nu^{r}} \phi\big(\nod{r}{0},\nod{k}{\n}\big)\,\apk{r}{0}  \right],
\end{equation}
where $c$ is the sum of all the weighted paths from node $\nod{r}{p}$ to node $\nod{k}{\n}$ ($p\leqslant \n, r \leqslant k$). We have the following lemma.
\begin{lemma}
The weighted sum $\phi\big(\nod{r}{p},\nod{k}{\n}\big)$ verifies
\begin{equation*}
	\phi(\nod{k}{\n},\nod{k}{\n}) = 1,
\end{equation*}
 and verifies also the following recurrence formulas:
\begin{equation}
 \label{pathnod}
 \begin{aligned}
 \phi\big(\nod{0}{p},\nod{k+1}{\n}\big) &= \sum\limits_{q=p}^{\n}\phi\big(\nod{0}{p},\nod{k}{q}\big)\Anp{q}{\n} + \sum\limits_{q=p}^{\n-1}\phi\big(\nod{0}{p},\nod{k+1}{q}\big)\Bnp{q}{\n},& 1\leqslant p\leqslant \n,\\
\phi\big(\nod{r}{0},\nod{k+1}{\n}\big) &=\sum\limits_{q=1}^{\n}\phi\big(\nod{r}{0},\nod{k}{q}\big)\Anp{q}{\n} + \sum\limits_{q=1}^{\n-1}\phi\big(\nod{r}{0},\nod{k+1}{q}\big)\Bnp{q}{\n},& 0\leqslant r< k,\\
\phi\big(\nod{k}{0},\nod{k+1}{\n}\big) &= \sum\limits_{q=0}^{\n}\phi\big(\nod{k}{0},\nod{k}{q}\big)\Anp{q}{\n} + \sum\limits_{q=1}^{\n-1}\phi\big(\nod{k}{0},\nod{k+1}{q}\big)\Bnp{q}{\n},\\
\phi\big(\nod{k+1}{0},\nod{k+1}{\n}\big) &= \sum\limits_{q=0}^{\n-1}\phi\big(\nod{k+1}{0},\nod{k+1}{q}\big)\Bnp{q}{\n} .
 \end{aligned}
\end{equation}
\end{lemma}
The proof of these equalities can be verified via the graph in \cref{fig1_graphnk}. Having them, we can state the following theorem.
\begin{theorem}
	\label{theoremakn}
If $\X_0(\bx) = \bu(0,\bx)$, thus $\apk{0}{0}=1$, $\apk{k}{0}=0,\,k\geqslant 1$ and $\apk{0}{p} = 0, p\geqslant 1$. We have also:
\begin{equation}
	\label{gen_formula_akn}
 \apk{k}{\n} = \frac{\nu^k}{k!}  \phi\big(\nod{0}{0},\nod{k}{\n}\big)\, \quad k\geqslant 1.
\end{equation}
\end{theorem}

If one can find the behaviour of $\Anp{p}{\n}$ and $\Bnp{p}{\n}$, \cref{apkpgdfnk} will be useful for asymptotic behaviour of $\apk{k}{\n}$. Both weights depend on $\X_\n$, which will be found in the next section.

\subsection{Finding $\X_\n$}
\label{sec_Xn}
Using the same strategy, we multiply \cref{diff-pgd} by $\T_\n$ and integrate the equation over the time interval $[0,\rangeT]$ to obtain the following equation governing $\X_\n$:
\begin{equation}
\label{eq_Xn}
 \sum\limits_{p=0}^{\n} \ppk{p}{\n} \X_p = \nu\sum\limits_{p=0}^{\n}\npk{p}{\n} \Delta \X_p,
\end{equation}
where
\begin{align}
\label{coeffnpkppktn}
 \npk{p}{\n} & = \int_0^\rangeT \T_p(t) \cdot \T_\n(t)\,\d t, & \ppk{p}{\n} = \int_0^\rangeT \T_p^{\prime}(t)\cdot \T_\n(t)\,\d t, \quad \forall p \leqslant \n
\end{align}
By arranging terms in \cref{eq_Xn}, the equation governing $\X_\n(\bx)$ is given below:
\begin{equation}
\label{eq_Xn0}
  \ppk{\n}{\n} \X_\n(\bx) - \nu\npk{\n}{\n} \Delta \X_\n(\bx) =  \sum\limits_{p=0}^{\n-1} \left( \nu \npk{p}{\n} \Delta \X_p(\bx) - \ppk{p}{\n} \X_p(\bx) \right),
\end{equation}
which could be written, if $\ppk{n}{\n}\neq 0$, in a compact form as follows:
\begin{equation}
 \label{eq_Xn_compact1}
  \left[\I - \nu\epk{\n}{\n} \Delta \right]\X_\n(\bx)  = -\sum\limits_{p=0}^{\n-1}  \gpk{p}{\n} \,\left[\I - \nu\epk{p}{\n} \Delta \right] \X_p(\bx),
\end{equation}
with $\I$ is the identity operator,  and
\begin{align}
\label{coeffepkgpk}
\epk{p}{\n} &= \frac{\npk{p}{\n}}{\ppk{p}{\n}}, &
\gpk{p}{\n} &= \frac{\ppk{p}{\n}}{\ppk{\n}{\n}}.
\end{align}
By arranging terms, let us denote by $\Fpgd(p,\n) =  \X_p - \nu\vepk{p}{\n} \Delta \X_{p-1}$. Having $\T_0(t)=1$ thus $\ppk{0}{\n} = 0$, then \cref{eq_Xn_compact1} is written in the recurrence formula:
\begin{equation}
 \label{Xn_fn}
 \Fpgd(\n,\n) = \Gpgd(\n,\n) + \sum\limits_{p=1}^{\n-1}  \Fpgd(p,\n) \,\wnp{p}{\n},
\end{equation}
with:
\begin{align*}
 \Gpgd(p,\n) &= \nu\epk{p}{\n} \Delta \X_p,&  \wnp{p}{\n} &= -\gpk{p}{\n}, & \vepk{p}{\n}&= \frac{\npk{p-1}{\n}}{\ppk{p}{\n}}.
\end{align*}
This recurrence formula is reduced using \cref{formula_fn} to obtain:
\begin{equation}
 \label{eq_Fn_compact}
 \Fpgd(\n,\n) = \sum\limits_{p=1}^{\n}\psi(p,\n)\, \Gpgd(p,\n),
\end{equation}
with $\psi(p,\n)$ to be given by \cref{formula_psi}.
By arranging terms, we can obtain the recurrence formula generating $\X_\n$:
\begin{equation}
 \label{Xn_fn1}
 \begin{aligned}
  \G_{\n}\cdot\X_\n &= \left[\nu\vepk{\n}{\n}\Delta \X_{\n-1} +  \sum\limits_{p=1}^{\n-1}  \eta_{p}^{\n}\,\Delta \X_p \right],
 \end{aligned}
\end{equation}
which simplifies the complexity of \cref{eq_Xn_compact1}, with:
\begin{align}
\G_{\n} &=\left[\I - \epk{\n}{\n} \Delta \right],& \eta_{p}^{\n} &= \nu\epk{p}{\n}\,\psi(p,\n).
\end{align}

To this end, finding $\X_\n$ and $\T_\n$ are obtained within the Picard iterative process: $\X_\n^0$ and $\T_\n^0$ are initiated and for $j>0$ $\X_\n^{j+1}$ is computed using $\T_\n^{j}$, then $\T_\n^{j+1}$ is found using the last $\X_\n^{j+1}$. Thus, two sequences are constructed, leading to the study of the asymptotic behaviour of $\apk{k}{\n}$ and $\X_\n$.

\subsection{From \ac{PGD} to \ac{TSE}}
If both formulas, \cref{u_tse} and \cref{u_pgd}, are solutions to the diffusion equation, are they equal? Formal calculus shows the equality. To check this, we replace $\T_n(t)$ by its time series representation in \cref{Tp_tse} to find:
\begin{equation}
 \label{equt1}
 \pu(t,\bx) = \sum\limits_{\n=0}^{\infty} \sum\limits_{k=0}^{\infty} \apk{k}{\n}\, t^k \X_\n(\bx).
\end{equation}
If the conditions are valid to permute the $\sum\limits_{\n=0}\sum\limits_{k=0}$, we can write the above equation as follows:
\begin{equation}
 \label{equt2}
 \pu(t,\bx) = \sum\limits_{k=0}^{\infty} \left(\sum\limits_{\n=0}^{\infty}  \apk{k}{\n} \X_\n(\bx)\right)\cdot t^k =  \sum\limits_{k=0}^{\infty} \huk_k(\bx) \,  t^k ,
 \end{equation}
which gives us the time series expansion of the \ac{PGD} solution where:
\begin{equation}
 \label{uk_by_pgd}
\huk_k(\bx) \coloneqq \sum\limits_{\n=0}^{\infty}  \apk{k}{\n} \X_\n(\bx) \stackrel{\text{\cref{theoremakn}}}{=} \frac{\nu^k}{k!} \sum\limits_{\n=1}^{\infty}     \phi\big(\nod{0}{0},\nod{k}{\n}\big) \X_\n(\bx), \quad k\geqslant 1.
\end{equation}
The above definition in \cref{uk_by_pgd} verifies formally \cref{uk} when $\n$ tends to $\infty$: By replacing $\T_\n(t)$ by its \ac{TSE} formula in \cref{diff-eq} and permuting the sum between $k$ and $\n$ we can show that:
\begin{equation}
 \begin{aligned}
\huk_{k+1}(\bx) = \sum\limits_{\n=0}^{\infty}  \apk{k+1}{\n} \X_\n(\bx) &= \frac{\nu}{k+1}\sum\limits_{\n=0}^{\infty}  \apk{k}{\n} \Delta\X_\n(\bx) = \frac{\nu}{k+1} \Delta \huk_k(\bx).
 \end{aligned}
\end{equation}
This equality is formal, subject to how tensor elements, $\X_\n$ and $\T_\n$, are computed, being orthogonal to the residual of the equation. \cref{uk_by_pgd} can be used to approximate $\huk_k$ as follows:
\begin{equation}
 \label{uk_app_by_pgd}
 \huk_k(\bx) \approx   \frac{\nu^k}{k!}\sum\limits_{\n=1}^{\rankm_k} \phi\big(\nod{0}{0},\nod{k}{\n}\big) \X_\n(\bx) \quad k\geqslant 1,
\end{equation}
where $\rankm_k$ is a rank to be chosen. This approximation is beneficial to generate new series terms $\huk_k$ as they will be a combination of already computed terms $\X_\n$ and the sum of weights $\phi$, depending also on $\X_\n$, which removes the need for the Picard iterative process, but becomes more complex with  $\phi$.

Next, will present an alternative method to compute $\mX_\n(\bx)$ in the context of \ac{PGD} by considering $\T_\n(t) = t^\n$. This proposal will ease the computation and establish the coefficients of the stabilization technique in the case of \ac{TSE}.

\subsection{\acf{SPGD}}

In the following, we consider the case where $\T_\n(t) = t^\n$. We denote by the process arising in this case the \acf{SPGD}. This simplified method finds $\mX_\n$ up to a finite rank, such that the residual is orthogonal to $t^\n$. It aims to skip the iterative process in finding $\X_\n$ and $\T_\n$. The solution is to be approximated by:
\begin{equation}
 \label{third_Alt}
 \spu(t,\bx) = \sum\limits_{\n=0}^{\infty} \mX_\n(\bx)\,t^\n.
\end{equation}
Thus, following the procedure in \cref{sec_Xn}, the coefficients in  \cref{coeffnpkppktn} are obtained as follows:
\begin{align}
 \label{npktn}
 \npk{p}{\n} &= \frac{\rangeT^{p+\n+1}}{p+\n+1},& \ppk{p}{\n} &= \frac{p
 \rangeT^{p+\n}}{p+\n}.
 \frac{2p\rangeT^{p-\n}}{p+\n}.
\end{align}
We go back to \cref{eq_Xn} and replace $\ppk{p}{\n}$ and $\npk{p}{\n}$, knowing that $\ppk{0}{\n} = 0$. We divide by $\cfrac{\rangeT^{2\n}}{2\n}$ on both sides and, manipulating terms, we obtain the following relation:
\begin{equation}
 \label{eq_Xn_update}
 \n\mX_\n - \nu\Delta \mX_{\n-1} -\frac{2\n \nu\rangeT}{2\n+1} \Delta \mX_\n + \sum\limits_{p=1}^{\n-1} \frac{2\n}{(p+\n)\rangeT^{\n-p}}\Big(p\mX_p - \nu\Delta \mX_{p-1}\Big) = 0,
\end{equation}
subject to the following orthogonality condition:
\begin{equation*}
-2\n \sum\limits_{p = \n+1}^{\infty} \Big( \frac{p}{\n+p} \mX_p - \frac{\nu \rangeT}{\n+p+1} \Delta \mX_p \Big) \cdot \rangeT^{p-\n} = 0.
\end{equation*}

For $\n=1$, we can easily obtain the equation that governs $\mX_1$ as follows:
\begin{equation*}
 \mX_1 - \frac{2\nu \rangeT}{3} \Delta \mX_1 = \nu\Delta \mX_0,
\end{equation*}
which resembles to \cref{stab_uk} that is used to compute a stabilized $\uk_1$ with a coefficient equal to $\staba_0 = {2\nu\rangeT}/{3}$. Computing $\mX_2$ involves $\mX_1 - \nu\Delta\mX_0$, which is not zero in this case. If we consider it zero, we will retrieve the case of computing $\uk_k$ without stabilization. Thus, the \ac{PGD}, when considering $\T_\n(t) = t^\n$, can be seen as computing $\uk_\n$ with stabilization technique.

Using graph theory, and denoting by $\Falt(\n) = \n \mX_\n - \nu\Delta \mX_{\n-1}$,  \cref{eq_Xn_update} can be written as follows:
\begin{equation}
 \label{eq_Xn_fn}
 \Falt(\n) =   \Galt(\n) + \sum\limits_{p=1}^{\n-1}  \Falt(p) \,\wnp{p}{\n},
\end{equation}
with
\begin{align}
\Galt(p) & = \stab_p \Delta \mX_p,&
 \stab_\n &= \frac{2\n \nu \rangeT}{2\n+1},&
 \wnp{p}{\n} & = \frac{-2\n}{(p+\n)\rangeT^{\n-p}},
\end{align}
Thus, we can prove that $\mX_\n$ is governed by the following equation:
\begin{equation}
 \label{eq_Xn_final}
 \Falt(\n) = \sum\limits_{p=1}^{\n}\psi(p,\n)\,\Galt(p),
\end{equation}
where coefficients $ \bnp{p}{\n}\coloneqq \psi(p,\n)$ are given by formula \eqref{formula_psi} as follows:
\begin{align}
\label{bnp}
\bnp{p}{\n} &= \sum\limits_{P\in \P(p,\n)} \prod\limits_{(i,j)\in P} \wnp{i}{j}.
\end{align}
The coefficient $\bnp{p}{\n}$ represents a contribution product for a simple path from $p$ to $\n$ in a direct graph where edges are labelled by weights $\wnp{i}{j}$.
Hereafter, we present the equations that govern the first three terms $\mX_p$ for $p=1,2,3$:
\begin{equation}
 \begin{aligned}
  \mX_1 - \stab_1 \Delta \mX_1 &= \nu\Delta \mX_0,\\
  2\mX_2 -\stab_2 \Delta \mX_2 &= \nu\Delta \mX_1 +  \stab_1\wnp{1}{2}  \Delta \mX_1,\\
  3\mX_3 -\stab_3 \Delta \mX_3 & =\nu\Delta \mX_2 +\stab_2\wnp{2}{3}\Delta \mX_2  + \stab_1 (\wnp{1}{3} + \wnp{1}{2}\wnp{2}{3})\Delta \mX_1.
 \end{aligned}
\end{equation}
 \cref{Alg_Xn_diff} presents the process for any $\n$. After that, the solution is approximated by:
\begin{equation}
 \label{}
 \sum\limits_{p=0}^{\n} \mX_p t^p,\quad \forall t\in [0,\rangeT].
\end{equation}
Note that $\rangeT$ represents the time step associated with this numerical flow.
\begin{algorithm}[H]
\caption{Compute $\mX_\n$ Sequence.\label{Alg_Xn_diff}}
\begin{algorithmic}[1]
\Require $\mX_0 = \bu(0,\bx)$; $\n$, integer; $\stab_p$, $\wnp{p}{\n}$, array
\For{$p \gets 1$ to $\n$}
    \State $A_p \gets p \,\I - \stab_p \Delta $
    \State $B_p \gets \text{Compute Path Sum}(p,\wnp{j}{p})$
    \State $L_p \gets \nu \Delta \mX_{p-1}$
    \For{$j \gets 1$ to $p-1$}
        \State $L_p \gets L_p + \stab_p\cdot B_p[j] \Delta\mX_j$
    \EndFor
    \State $X_p \gets A_p^{-1}\circ L_p$
\EndFor
\State \Return $X_n$
\end{algorithmic}
\end{algorithm}

\subsection{Summary}
In the case of the diffusion equation, if one can write the approximation  as follows:
\begin{equation}
	\sum\limits_{\n=0}^{\infty}\X_\n(\bx)\T_\n(t),
\end{equation}
and one can formulate the equation finding $\X_\n$ depending on the last $\X_p$ as follows:
\begin{equation}
	\label{gen_form_Xn}
	\X_\n = \left[ \I - \lambda_\n\Delta \right]^{-1} \circ \left[\ \Theta_{\n}\Delta \X_{\n-1} + \sum\limits_{p=1}^{\n-1} \theta_p^\n \Delta \X_p  \right],
\end{equation}
thus the following table describe the coefficients of stabilization, $\lambda_\n$ ,$\Theta_{\n}$ and $\theta_p^\n$ for every method:

\begin{table}[!ht]
\begin{center}
\caption{ List of the methods and their coefficients aligned with \cref{gen_form_Xn}. \label{tab1}}
	\renewcommand{\arraystretch}{2}
	\begin{tabular}{|c|c|c|c|c|}
		\hline
		\multirow{2}{*}{\ac{TSE}}& $\X_\n = \uk_\n$  & \multirow{2}{*}{$\lambda_\n = 0$} &  $\Theta_{\n} = {\nu}/{\n}$\\
								 & $\T_\n(t) = t^\n$  & & $\theta_p^\n =0$\\
		\hline
		\multirow{2}{*}{\ac{STSE}}& $\X_\n = \uk_\n$  & \multirow{2}{*}{$\lambda_\n = \staba_{\n-1}$} & $\Theta_{\n} = {\nu}/{\n}$\\
		& $\T_\n(t) = t^\n$  & & $\theta_p^\n =0$\\
		\hline
		\multirow{2}{*}{\ac{SPGD}} & $\X_\n = \mX_\n$  & \multirow{2}{*}{$\lambda_\n = \stab_\n/\n$} & $\Theta_{\n} = {\nu}/{\n}$\\
		& $\T_\n(t) = t^\n$  & & $\theta_p^\n =\bnp{p}{\n}\stab_p/\n$\\
		\hline
		\multirow{3}{*}{\ac{PGD}} & \multirow{2}{*}{$\X_\n$} & \multirow{2}{*}{$\lambda_\n = \epk{\n}{\n}$} &  $\Theta_{\n} = \nu\vepk{\n}{\n}$  \\
		& & &  $\theta_p^\n  = \eta_{p}^{\n}$ \\
		& $\T_\n(t) = \sum\limits_{k=0}^{\infty} \apk{k}{\n}t^\n$  &$\apk{k}{\n} = \frac{\nu^k}{k!}  \phi\big(\nod{0}{0},\nod{k}{\n}\big)$ & \\
		\hline
	\end{tabular}
\end{center}
\end{table}

Throughout \cref{tab1}, we can see the connection between \ac{TSE} and \ac{PGD} passing by \ac{STSE} and \ac{SPGD}. Next, we will compare the computational efficiency of the stabilized \ac{TSE} method \ac{STSE} and the simplified \ac{PGD} method \ac{SPGD}.

\subsection{Numerical performance and computational efficiency}
In this section, we compare the efficiency of the numerical simulation performed using the \ac{STSE} and the \ac{SPGD}. The \ac{STSE} coefficients are varied to assess their performance. The numerical simulation is performed with the initial condition $\bu(0,\bx) = \sin(\pi\bx)$ for $\bx \in [0,1]$. The simulation is done for a set of values of time steps $\rangeT$ to reach the final instant $\mT= \N\cdot\rangeT$, where $\N$ is the number of time steps.

\begin{figure}[h!]
\centering
\includegraphics[scale=0.4]{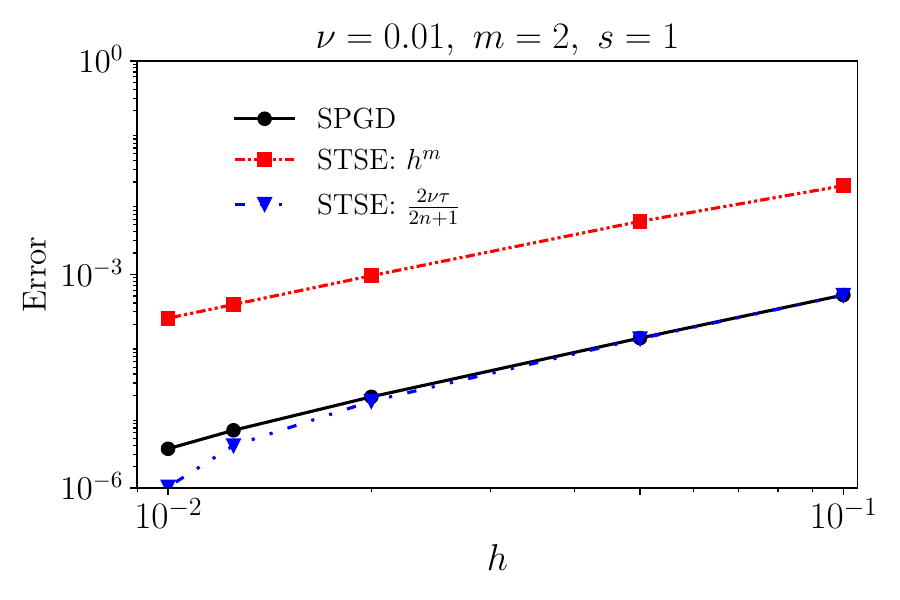}
\includegraphics[scale=0.4]{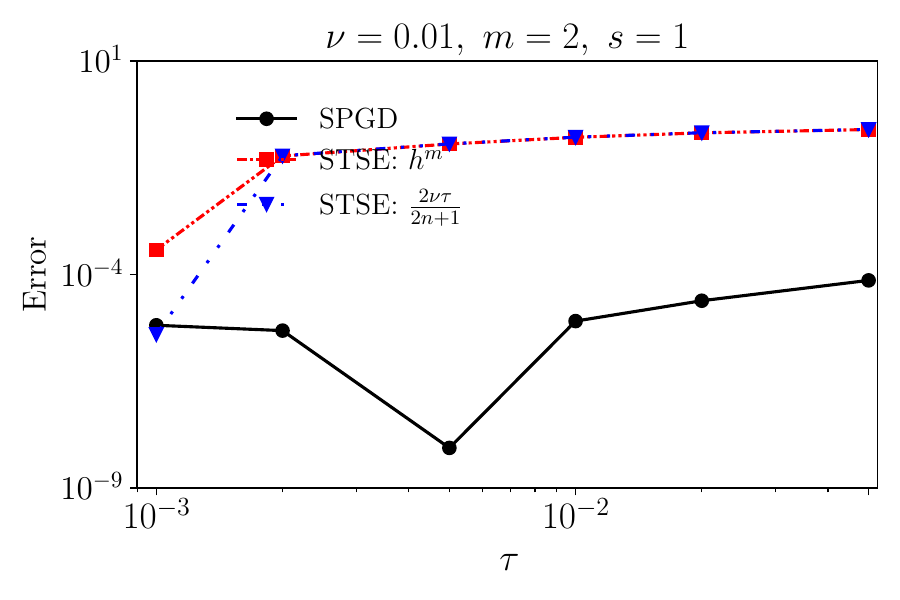}
\includegraphics[scale=0.4]{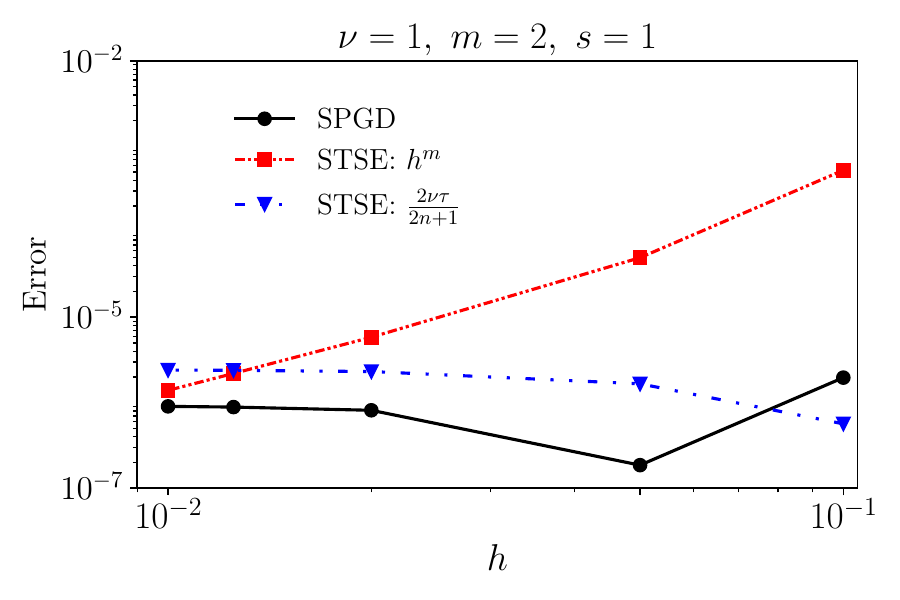}
\includegraphics[scale=0.4]{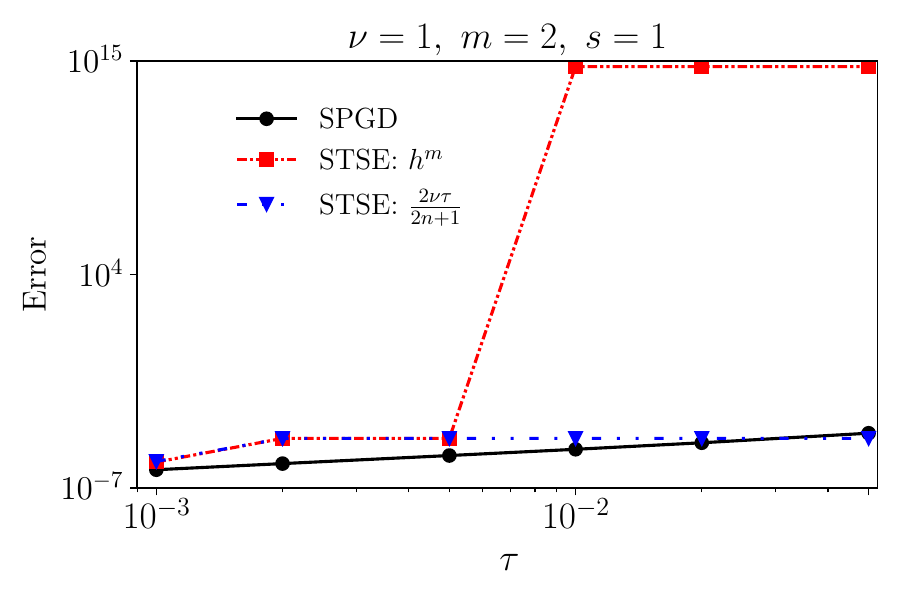}
\caption{Comparison of the error between \ac{SPGD}, \ac{STSE} with $\staba_0 = h^m$ stabilization, and \ac{STSE} with $\lambda_\n = \frac{2\nu\rangeT}{2\n+1}$ stabilization. Left column: error versus mesh size $h$. Right column: error versus time step $\rangeT$. Top row: diffusion coefficient $\nu = 0.01$. Bottom row: diffusion coefficient $\nu = 1$. All simulations use first-order finite elements ($s=1$).\label{fig:error_comparison}}
\end{figure}

Figure \ref{fig:error_comparison} presents a comparison of the error between the \ac{SPGD} method and two \ac{STSE} variants: one using stabilization coefficients of the form $\lambda_\n = c^\n\,h^m$, and the other using $\lambda_\n = \frac{2\nu\tau}{2n+1}$. The results are shown concerning the mesh size $h$ (left column) and the time step $\rangeT$ (right column), for two diffusion coefficients: $\nu = 0.01$ (top row) and $\nu = 1$ (bottom row). All simulations use a first-order finite element method ($s=1$).

For $\nu = 0.01$, the error decreases with decreasing $h$ for all methods, although \ac{SPGD} consistently achieves the lowest error. In contrast, when $\nu = 1$, the error remains nearly constant for \ac{SPGD} and the \ac{STSE} variant with $\stab_\n$ stabilization, while the \ac{STSE} using $h^m$ continues to exhibit decreasing error. This suggests that $h^m$-based stabilization may improve spatial convergence for higher diffusivity.

Regarding temporal accuracy, the \ac{SPGD} method shows insensitivity to the time step $\rangeT$ for both values of $\nu$, maintaining stable error across the range. However, for $\nu = 0.01$, the error of both \ac{STSE} variants increases with larger $\rangeT$, indicating sensitivity to time-step size. When $\nu = 1$, the \ac{STSE} variant with $\stab_\n$ coefficients exhibits superior stability, while the error in the $h^m$-based \ac{STSE} increases rapidly for larger time steps.

These results demonstrate that \ac{SPGD} is robust concerning spatial and temporal discretization, while the choice of stabilization in \ac{STSE} significantly affects performance, particularly in high-viscosity regimes.

\begin{figure}[h!]
\centering
\includegraphics[scale=0.4]{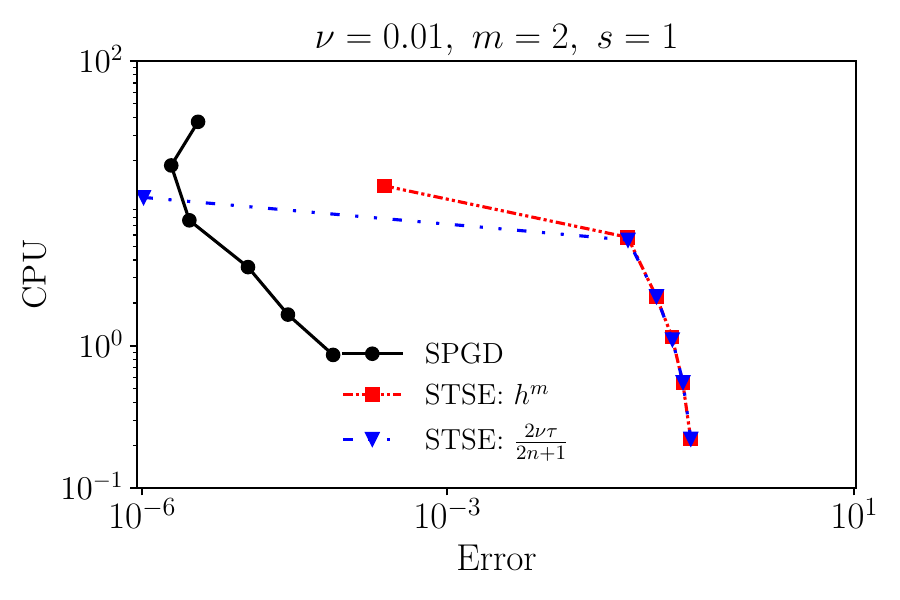}
\includegraphics[scale=0.4]{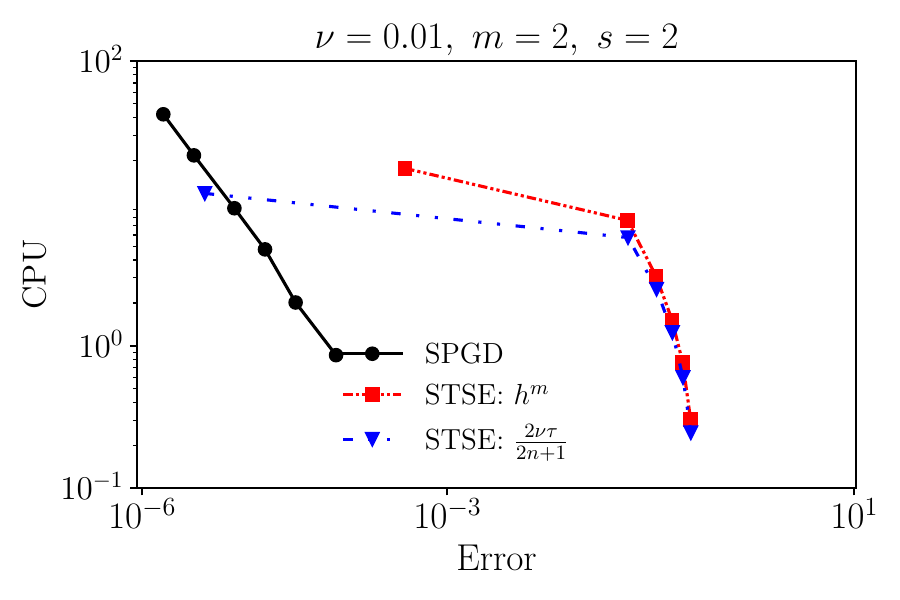}
\includegraphics[scale=0.4]{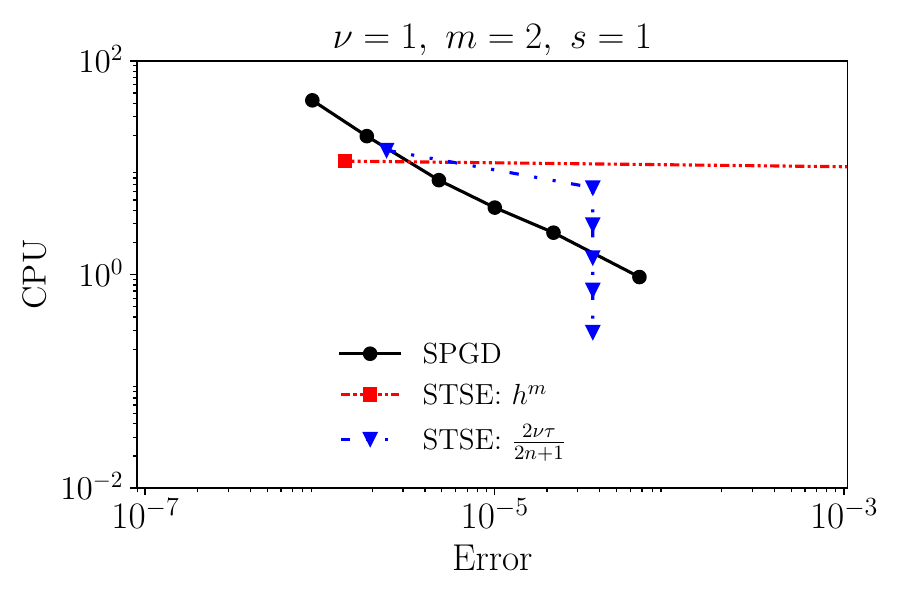}
\includegraphics[scale=0.4]{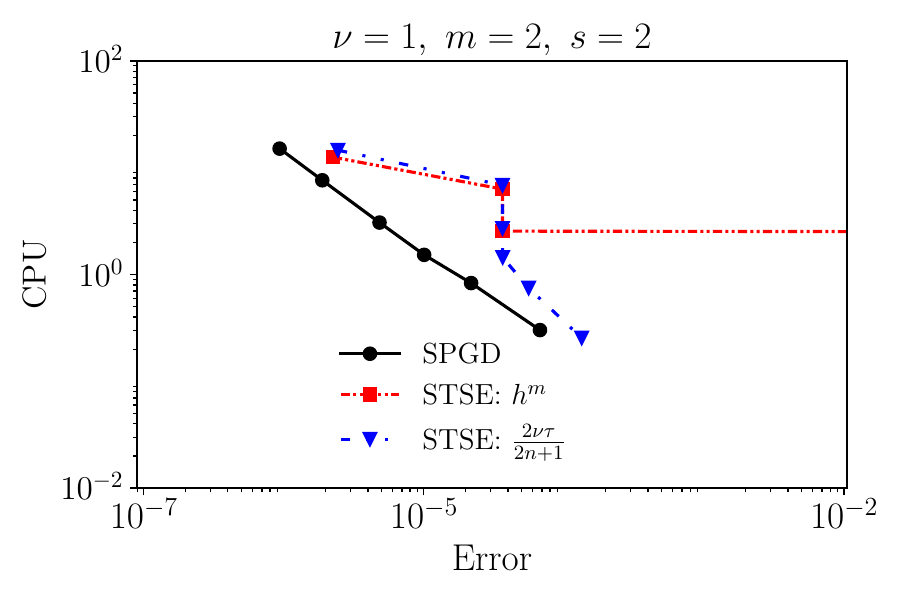}
\caption{CPU time versus error for \ac{SPGD}, \ac{STSE} with $h^m$ stabilization, and \ac{STSE} with $\lambda_\n = \frac{2\nu\tau}{2\n+1}$ stabilization. Left column: first-order finite elements ($s=1$); right column: second-order finite elements ($s=2$). Top row: diffusion coefficient $\nu = 0.01$; bottom row: $\nu = 1$. Simulations are performed with fixed mesh size $h = 0.01$ while varying the time step $\rangeT$. \label{fig:cpu_vs_error}}
\end{figure}

Figure \ref{fig:cpu_vs_error} illustrates the computational cost (CPU time) versus error for the \ac{SPGD} and \ac{STSE} methods using two different stabilization strategies: $\lambda_\n =h^m$ and $\lambda_\n = \frac{2\nu\tau}{2\n+1}$. The results are shown for two finite element orders: $s=1$ (left column) and $s=2$ (right column), and two diffusion coefficients: $\nu = 0.01$ (top row) and $\nu = 1$ (bottom row). The simulations were performed with a fixed mesh size $h = 0.01$ while varying the time step $\tau$. For each $\tau$, the total CPU time and corresponding error were recorded.

Across all cases, the \ac{SPGD} method exhibits a clear trade-off: as the time step decreases, the error improves but at the cost of increasing CPU time. This trend is expected due to the higher number of iterations required with finer temporal resolution.

In contrast, the \ac{STSE} method shows distinct behaviours depending on the stabilization strategy. The $h^m$-based \ac{STSE} variant suffers from severe error growth at larger time steps for the higher-viscosity regime, leading to inefficient performance. On the other hand, the \ac{STSE} variant using the $\stab_\n$ coefficients demonstrates greater stability: the error remains almost constant across different $\rangeT$ values, and the CPU time increases more slowly than \ac{SPGD}. This performance suggests that $\stab_\n$-based stabilization leads to a more robust method in terms of computational efficiency, especially in regimes where reducing the time step does not necessarily yield better accuracy.

Overall, \ac{SPGD} maintains the best accuracy-to-cost balance for high-precision requirements, while \ac{STSE} with $\stab_\n$ offers a more stable and cost-effective alternative for moderate accuracy thresholds.

\section{Application to Inviscid Flow}
\label{sec40}
We have seen before how the \ac{TSE} method connects to \ac{PGD} via graph theory. We have used this connection to find the coefficients of stabilization when employing an artificial diffusion technique, which also appears in the \ac{PGD} method for the presence of the Laplace term. However, what happens when the equation does not involve the diffusion term? How should the \ac{TSE} method be stabilized if needed? How is the graph that connects it to \ac{PGD} in this case? These questions will be investigated in the following section via the application to the Inviscid flow.

We consider the dimensionless system modelling the dynamics of inviscid flow:
\begin{equation}
\label{eq_inviscid}
 \left\lbrace
 \begin{aligned}
 \rho \left(\bu_t + (\bu\cdot\nabla) \bu\right) + \nabla \bp &= 0,\\
 \nabla \cdot \bu &= 0,
 \end{aligned}
 \right.
\end{equation}
where $\rho$ is the density of the fluid considered constant here, $\bu = (u_x,u_y)$ is the dimensionless velocity of the flow and $\bp$ is the dimensionless scalar pressure, $U_\infty$ and $P_\infty$ are the velocity and the pressure far field. For simplicity, we consider $P_\infty = U_\infty^2\rho$.

\subsection{Classical \ac{PGD} method}
The \ac{PGD} method consists of approximating the spacetime solution as follows:
\begin{align*}
 \bu(\bx,t) &= \sum\limits_{p=0}^{\n} \Yu_p(\bx) \T_p(t), & \bp(\bx,t) & =  \sum\limits_{p=0}^{\n} \Yp_p(\bx) \T_p(t),
\end{align*}
where $( \Yu_p(\bx), \Yp_p(\bx))$ and $ \T_p(t)$ are obtained using the Picard iterative process.

Providing numerical solutions to such problems is challenging for several reasons, where treating the slip-type boundary conditions ($\bu\cdot n = 0$), especially in \ac{FSI} problems. It could be implemented weakly using a Nitsche penalty \cite{Nitsche19719, ARAYA2024117037} on the normal velocity component, while tangential velocity remains free. It consists of adding
\begin{equation*}
 \frac{\eta}{h^2}\int_\Gamma (\bu\cdot \bn)^2
\end{equation*}
in the energy formulation of the problem, where $\eta$ is a scaling coefficient, $h$ is related to the mesh size and $\Gamma$ is a specified boundary element. However, the linear system governing the mixed formulation $(\bu,\bp)$ is asymmetric and generates instabilities. Another way to apply the boundary conditions is to use the so-called \ac{RT} Finite element for the velocity, which satisfies intrinsically the slip condition but makes applying Dirichlet boundary conditions difficult.

Another problem is the dealing with the convective term $(\bu.\cdot\nabla)\bu$ when computing the space and time modes. First, let us give the general form of computing the $\n^{th}$ space mode. It follows writing the approximation of $\bu$ and $\bp$ up to rank $\n$, injecting them in \cref{eq_inviscid}, multiplying by $\T_\n(t)$, and integrating over the time domain. This leads to the following equation, completed by  the continuity one, which governs the couple $( \Yu_\n(\bx), \Yp_\n(\bx))$
\begin{equation*}
\sum\limits_{p=0}^{\n} \left[\ppk{p}{\n}\Yu_p  + \sum\limits_{r=0}^{\n}\cpk{p,r}{\n}(\Yu_p\cdot \nabla)\Yu_r   + \npk{p}{\n}\nabla\Yp_p  \right] = 0,
\end{equation*}
with $\ppk{p}{\n}$ and $ \npk{p}{\n}$ defined in \cref{coeffnpkppktn} and
\begin{equation}
 \label{cpef_cpk}
 \cpk{r,p}{\n} = \int_0^\rangeT \T_r(t)\cdot\T_p(t) \cdot \T_\n(t)\,\d t.
\end{equation}
For example, computing the first space mode $( \Yu_1(\bx), \Yp_1(\bx))$ could be written as follows ($\cpk{r,p}{\n}$ is symmetric):
\begin{equation*}
\begin{aligned}
 \ppk{1}{1}\Yu_1 &+\cpk{0,1}{1} \left[(\Yu_1\cdot \nabla)\Yu_0 + (\Yu_0\cdot \nabla)\Yu_1  \right] + \cpk{1,1}{1}(\Yu_1\cdot \nabla)\Yu_1 + \npk{1}{1}\nabla \Yp_1  =\\
 &  - \Yu_0 \ppk{0}{1} - \cpk{0,0}{1}(\Yu_0\cdot \nabla)\Yu_0 - \npk{0}{1}\nabla \Yp_0.
\end{aligned}
\end{equation*}
The non-linear term, multiplying $ \cpk{1,1}{1}$, could be dealt with during the Picard iterative process: having reaching the $i^{th}$ iteration with $( \Yui{i}_1,\, \Ypi{i}_1,\, \Ti{i}_1)$, one can solve the linearized system as follows:
\begin{equation*}
 \begin{aligned}
 \Yui{i+1}_1 + \left(\left[ \frac{\cpk{0,1}{1,i}}{\ppk{1}{1,i}} \Yu_0 +  \frac{\cpk{1,1}{1,i}}{\ppk{1}{1,i}} \Yui{i}_1\right]\cdot \nabla \right)\Yui{i+1}_1 + \frac{\npk{1}{1,i}}{\ppk{1}{1,i}}\nabla \Ypi{i+1}_1 =\\
 -\frac{\ppk{0}{1,i}}{\ppk{1}{1,i}} \Yu_0 -\left(\left[ \frac{\cpk{0,0}{1,i}}{\ppk{1}{1,i}} \Yu_0 +  \frac{\cpk{0,1}{1,i}}{\ppk{1}{1,i}} \Yui{i}_1\right]\cdot \nabla \right)\Yu_0 - \frac{\npk{0}{1,i}}{\ppk{1}{1,i}}\nabla \Yp_0.
\end{aligned}
\end{equation*}
The frozen velocity ($\bu^\star \coloneqq \frac{\cpk{0,1}{1,i}}{\ppk{1}{1,i}} \Yu_0 +  \frac{\cpk{1,1}{1,i}}{\ppk{1}{1,i}} \Yui{i}_1$) in the left hand side, which is to be updated at each step $i$, is not the same as the right one, contrary to the case when replacing the convective term with a skew-symmetric one \cite{MORINISHI199890}. In addition to that, coefficients $\cpk{\times,\times}{1,i},\, \npk{1}{1,i}$ and $\ppk{1}{1,i}$, evaluated with time modes $\Ti{i}_p(t)$ at the $i^{th}$ iteration, are updated in the frozen velocity and in the right hand side of the system. These are sources of oscillations in the Picard process.

When it comes to solving the equation governing the time mode, one might find the approximations on a set of discrete points $\{t_i\} \subset [0,T]$, and the discrete form of the equation can be assembled into a linear system as follows:
\[
\mathrm{A}_t \bT_\n = b_\n, \quad \mathrm{A}_t = \frac{\rho}{\Delta t} \mathrm{B}+ \rho \sigma_x \M_t, \quad \sigma_x = \int_\Omega N_{\text{skew}}(\bu^*;\X,\mathbf{v})\X,
\]
where $\A_t$ is the matrix containing the discrete form of the time derivative represented in $\mathrm{B}$ (such as the second order \ac{BDF}) and the time mass matrix $\M_t$, $\bT_\n$ is the vector representing the approximation at the nodes $\{t_i\}$. The choice of the operator $\mathrm{B}$ affects the linear solving and thus the Picard iteration: If one chooses the Backward-Euler method, the matrix $\mathrm{A}_t$ is close to Semi-Positive-Definite, but the approximation is of first order, while choosing \ac{BDF}2 will generate a non-symmetric matrix $\mathrm{A}_t$ and can cause oscillatory update of $\bT$, thus a regularisation (preconditioning) is to be applied. To overcome most of these difficulties, we will extend the connection between \ac{TSE} and \ac{PGD} technique to the inviscid flow equation.

\subsection{From \ac{TSE} to \acf{SPGD}}
By writing the solution of $\bu$ and $\bp$ in their \ac{TSE} forms:
\begin{align}
\label{u_p_TSE}
 \tu(t,\bx)& = \sum\limits_{k=0}^{\infty} \uk_k(x) t^k, & \tp(t,\bx)& =\sum\limits_{k=0}^{\infty} \pk_k(x) t^k,
\end{align}
We inject them in \cref{eq_inviscid} to find the recurrence formula governing $(\uk_k,\pk_{k-1})$ by equating terms of $t^k$. We find:
\begin{equation}
 \label{eq_invscid_tk}
 \left\lbrace
 \begin{aligned}
 \Fif(k) \coloneqq k \uk_{k} +\sum\limits_{r=0}^{k-1} (\uk_{r}\cdot\nabla)\uk_{k-r-1}  + \nabla \pk_{k-1}  & = 0,\\
 \nabla \cdot \uk_k &= 0.
 \end{aligned}
 \right.
\end{equation}
This system is to be solved in \ac{FEM} framework using the mixed formulation subject to the boundary conditions. The \ac{STSE} consists of adding an artificial diffusion, a regularization term, to the left-hand side:
\begin{equation}
 \label{eq_invscid_stse}
 \left\lbrace
 \begin{aligned}
 \Fif(k) +\alpha_k \Delta \uk_k & = 0,\\
 \nabla \cdot \uk_k &= 0.
 \end{aligned}
 \right.
\end{equation}
Next, we present the computation of modes via the \ac{SPGD}. For that, we consider approximating the solution of $\bu$ and $\bp$ as follows:
\begin{align}
 \label{u_p_alt}
 \spu(t,\bx) &= \sum\limits_{p=0}^{\infty} \Xu_p(\bx)\, t^p, &
 \spp(t,\bx) &= \sum\limits_{p=0}^{\infty} \Xp_p(\bx)\, t^p,
\end{align}
and finding $\Xu_p$ and $\Xp_p$ having $t^p$ orthogonal to the residual of the equation. For this problem, having only the initial condition of the velocity, namely $\bu(0,\bx)$, we consider that $\Xu_0(x) = \bu(0,\bx)$. Then we proceed to find $\Xu_1$ and $\Xp_{0}$ and continue until reaching the level $\n$ in finding $\Xu_{\n}$ and $\Xp_{\n-1}$. Before proceeding, let us write the non-linear (convective) term in the equation as follows:
\begin{equation*}
\begin{aligned}
 \left( \sum\limits_{p=0}^{\n} \Xu_p(\bx)\, t^p. \nabla \right) \sum\limits_{p=0}^{\n} \Xu_p(\bx)\, t^p &= \sum\limits_{p=0}^{\n} \left(\sum\limits_{r=0}^{p}(\Xu_r\cdot\nabla)\Xu_{p-r} \right)t^p \\
 &+ \sum\limits_{p=1}^{\n} \left(\sum\limits_{r=0}^{\n-p} (\Xu_{r+p}\cdot \nabla) \Xu_{\n-r} \right) t^{\n+p}.
 \end{aligned}
\end{equation*}
We replace the time derivative in its \ac{SPGD} form, the convective term and the pressure in \cref{eq_inviscid}, then we multiply by $t^\n$ and integrate over the time interval $[0,\rangeT]$, and divide by $\frac{\rangeT^{2\n}}{2\n}$ to find the following equation:
\begin{equation}
 \label{SPGD_inviscid}
 \begin{aligned}
  \sum\limits_{p=1}^{\n} \left( p\Xu_p  + \sum\limits_{r=0}^{p-1} \big( \Xu_r\cdot \nabla \big) \Xu_{p-1-r}  +\nabla \Xp_{p-1} \right) \frac{2\n}{(\n+p)\rangeT^{\n-p}}& \\
  +\sum\limits_{p=0}^{\n} \left( \sum\limits_{r=0}^{\n-p} \big(\Xu_{r+p}\cdot \nabla\big)\Xu_{\n-r} \right) \frac{2\n\rangeT^{p+1}}{2\n+p+1} &= 0,\\
\nabla\cdot \Xu_p&=0.
 \end{aligned}
\end{equation}
By denoting:
\begin{equation}
  \label{Hn_inviscid}
\begin{aligned}
 \Gif(\n) &\coloneqq - \sum\limits_{p=0}^{\n} \left( \sum\limits_{r=0}^{\n-p} \big(\Xu_{r+p}\cdot \nabla\big)\Xu_{\n-r} \right)\frac{2\n\rangeT^{p+1}}{2\n+p+1},
\end{aligned}
\end{equation}
we can express \cref{SPGD_inviscid} as follows:
\begin{equation}
\label{rec_if}
\Fif(\n) = \Gif(\n) + \sum\limits_{p=1}^{\n-1}\Fif(p)\, \wnp{p}{\n},\quad  \wnp{p}{\n}= \frac{-2\n}{(\n+p)\rangeT^{\n-p}},
\end{equation}
thus, by using the recurrence formula in \eqref{rec_fromula} we can simplify the formula:
\begin{equation}
 \Fif(\n) =\sum\limits_{p=1}^{\n} \psi(p,\n)\, \Gif(p).
\end{equation}
The right-hand side of the last equation is the additional term to be added to go from \ac{TSE} to \ac{SPGD}. It is written as follows for $\n=1$:
\begin{equation*}
 \Xu_1 + (\Xu_0\cdot\nabla)\Xu_0  + \nabla \Xp_0 = - \Big( (\Xu_0\cdot\nabla)\Xu_1 +  (\Xu_1\cdot\nabla)\Xu_0\Big)\frac{2\rangeT}{3} - \Big( (\Xu_1\cdot\nabla)\Xu_1\Big)\frac{\rangeT^2}{2}.
\end{equation*}
We can see that the left-hand side of the equation has the same structure as in \cref{eq_invscid_tk} for $k=1$, but the right-hand side contains two additional terms of order $\rangeT$ and $\rangeT^2$. We can neglect the term of $\rangeT^2$ for small $\rangeT$, which reduces the computational cost and skips a Picard iterative process. The term of $\rangeT$ might be added in the time series stabilization technique with the artificial diffusion term.

For $\n=2$, here what becomes \cref{rec_if} governing $(\Xu_2,\Xp_1)$:
\begin{equation}
 \label{Xu2_Xp1}
 \begin{aligned}
  2\Xu_2 + \sum\limits_{r=0}^{1}\left(\Xu_r\cdot\nabla\right)\Xu_{1-r} + \nabla \Xp_1& =\psi(1,2) \Gif(1) + \Gif(2),\\
  &= \frac{8}{9}\sum\limits_{r=0}^{1}\left(\Xu_r\cdot\nabla\right)\Xu_{1-r} + \frac{8\rangeT}{12} (\Xu_1\cdot\nabla)\Xu_1\\
   & \frac{-4\rangeT}{5} \left[(\Xu_{0}\cdot\nabla)\Xu_{2} + (\Xu_{1}\cdot\nabla)\Xu_{1}+ (\Xu_{2}\cdot\nabla)\Xu_{0}\right]\\
   & \frac{-4\rangeT^2}{6} \left[(\Xu_{1}\cdot\nabla)\Xu_{2} + (\Xu_{2}\cdot\nabla)\Xu_{1}\right]\\
&\frac{-4\rangeT^3}{7} \left[(\Xu_{2}\cdot\nabla)\Xu_{2}  \right].
 \end{aligned}
\end{equation}
While the left-hand side contains the same terms as for the \ac{TSE} for $k=2$, the right-hand side of the equation contains an additional term with $\Xu_2$, where the ones multiplying $\rangeT^3$ can be neglected for small $\rangeT$.

\subsection{Simulating the flow in a two-dimensional annulus.}

We consider the unsteady incompressible Euler equations in a two-dimensional annulus
$\Omega \subset \mathbb{R}^2$, representing the exterior of a circular cylinder of radius $a$
inside a far-field boundary of radius $R$. On the outer boundary $\Gamma_R$ ($r=R$), a time-dependent Dirichlet condition is imposed:
  \[
  \mathbf{u}(x,y,t) = (U(t),0),
  \]
where $U(t)$ is a smooth ramp reaching a constant inflow velocity $U_0=1$ after time $T=2$. On the cylinder surface $\Gamma_a$ ($r=a$), a slip condition is enforced:
  \[
  \mathbf{u}\cdot \mathbf{n} = 0,
  \]
which is implemented weakly using a Nitsche penalty on the normal velocity component, while tangential velocity remains free.
The pressure traction exerted by the fluid on
the cylinder is $-p\,\mathbf{n}$, so the net hydrodynamic force on the cylinder is
\[
\mathbf{F}(t) \;=\; \int_{\Gamma_a} -\,p(\mathbf{x},t)\,\mathbf{n}\,ds
\;=\; \left(
-\int_{\Gamma_a} p\,n_x\,ds,\;
-\int_{\Gamma_a} p\,n_y\,ds
\right)^\top.
\]

We run the simulation within the velocity-pressure mixed formulation in the \ac{FEM} \textit{FEniCS} framework and consider the case of the \ac{STSE} method for two stabilization coefficients. For the \ac{SPGD}, the function $\Gtse(\n)$ is also considered in the system of computing $(\Xu_\n, \Xp_{\n-1})$ for what the bigger time step to maintain the stability is to be taken of order $\rangeT \sim 10^{-4}$ in order to reach a stable simulation at $t=2$.

\begin{figure}[!ht]
\centering
\includegraphics[scale=0.5]{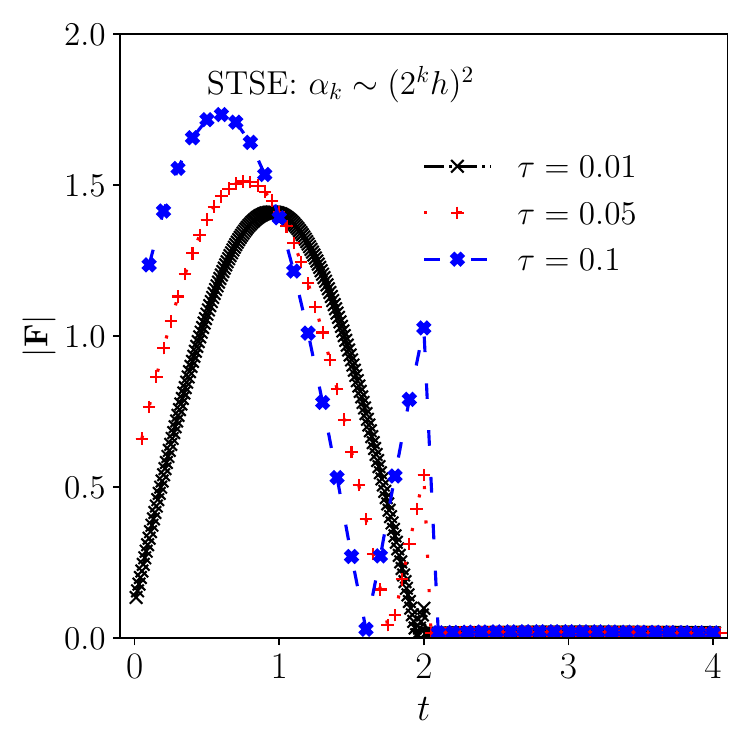}
\includegraphics[scale=0.5]{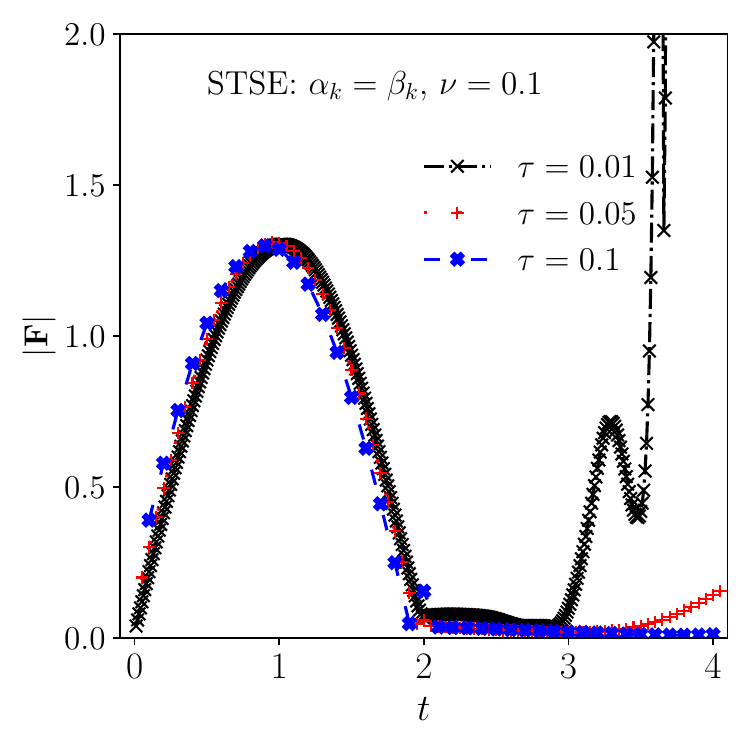}
\caption{The norm of the Force on the cylinder, both obtained with \ac{STSE}: In the left, the stabilization is considered of order $\staba_k\sim  (2^kh)^2$ and in the right with $\stab_k$ coefficients. This was done with different time steps $\rangeT$. \label{fig1_force_euler_stse}}
\end{figure}

We present in \cref{fig1_force_euler_stse} the norm of the Force $\mathbf{F}(t)$ on the cylinder obtained with the \ac{STSE} and with different time step $\rangeT$ over the time interval $[0,4]$. The plot on the left presents the Force considering the stabilization as of $\staba_k = (2^k h)^2$, while the right plot presents the simulation considering $\staba_k = \stab_k$.

Recall that $\stab_\n = \frac{2 \n \nu \rangeT}{2\n+1}$ and here we took $\nu = 0.1$. This explains why for small time steps, the simulation explodes at some instant $t$, while it remains stable for larger $\rangeT$. The stabilization coefficients become smaller. In this stabilization case ($\staba_k = \stab_k$), there is a trade-off between the precision and the stability if one considers a small damping coefficient. Nevertheless, this is not observed if considering the coefficients relative to the mesh size only ($\staba_k = (2^k h)^2$), but a loss of precision is observed around $t=0$ and $t=2$ for higher $\rangeT$.

\begin{figure}[!ht]
\centering
\includegraphics[scale=0.5]{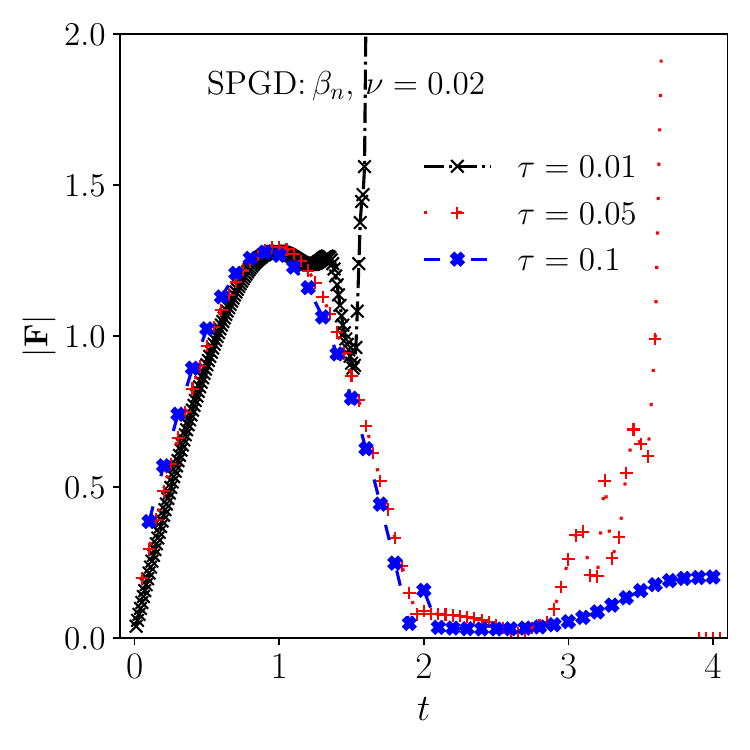}
\includegraphics[scale=0.5]{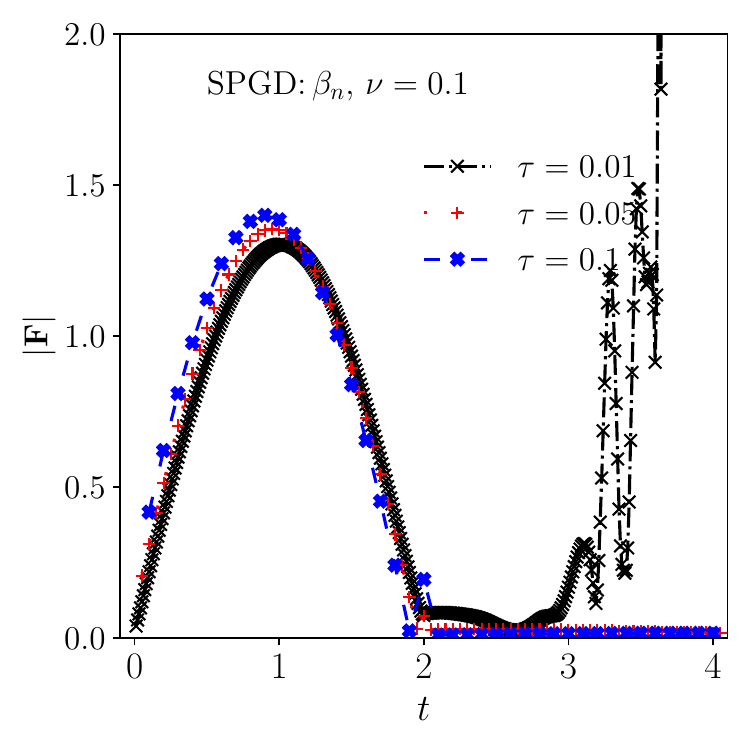}
\caption{The norm of the Force on the cylinder, both obtained with \ac{SPGD}: In the left, $\staba_\n$ is considered with $\nu=0.02$m, while on the right, $\nu=0.1$. This was done with different time steps $\rangeT$. \label{fig1_force_euler_spgd}}
\end{figure}

\begin{figure}[!ht]
 \centering
 \includegraphics[scale=0.35]{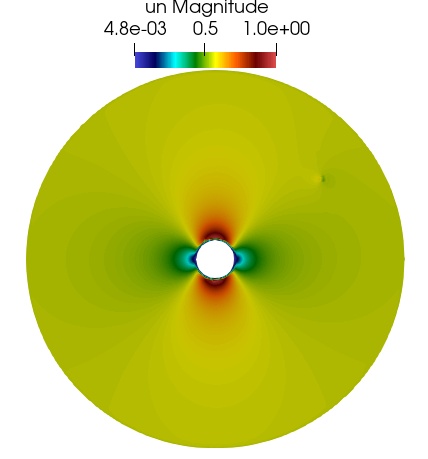}
 \includegraphics[scale=0.35]{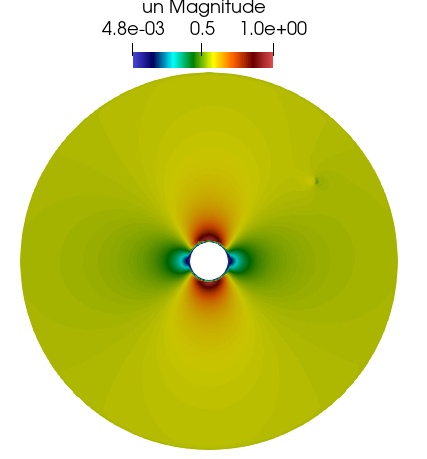}
 \includegraphics[scale=0.35]{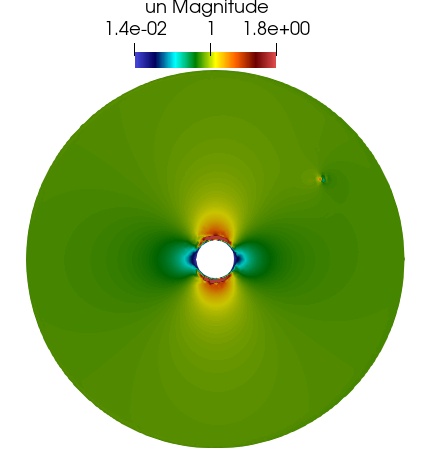}
 \includegraphics[scale=0.35]{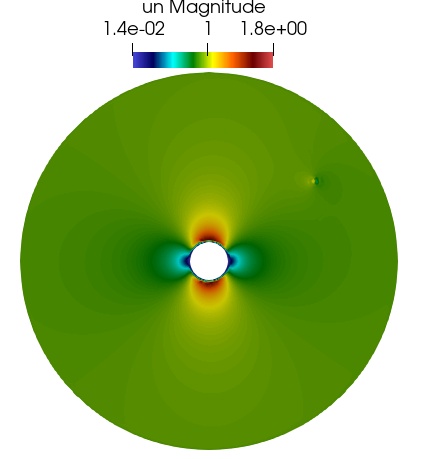}
 \includegraphics[scale=0.35]{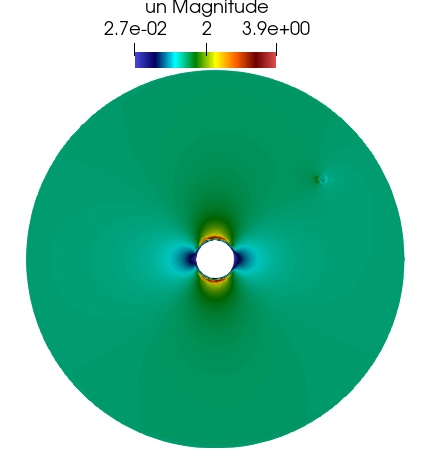}
 \includegraphics[scale=0.35]{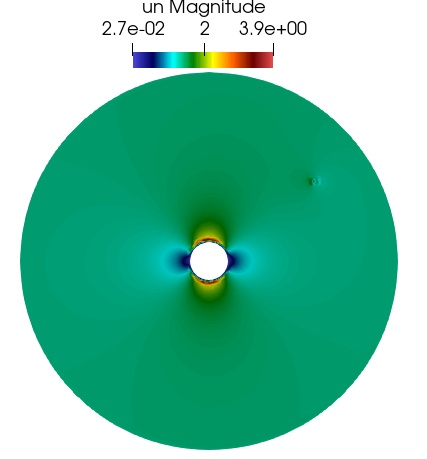}
\caption{The magnitude of the first, second, and third modes (from top to bottom) of the \ac{STSE} (left) and \ac{SPGD} (right) at $t=0.5$. \label{fig_euler_unt005}}
\end{figure}

When it comes to the \ac{SPGD} method, we consider two cases of $\stab_\n$ with $\nu=0.1$ and $\nu=0.02$ and plot in \cref{fig1_force_euler_spgd} the Force obtained. We remark that a smaller $\nu$ or $\rangeT$ implies an earlier simulation explosion for having a smaller damping effect in $\staba_\n$. This confirms the trade-off between precision and stability, which was also seen before in the case of the diffusion equation.
To this end, we present in \cref{fig_euler_unt005}  also the first three space modes $\Xu_p$ and $\Xp_p$ at different instants: $t=0.5$

\section{Application to Navier-Stokes equations}
\label{sec4}

In addition to extending the connection built above to the \ac{NS} equations, this section aims to provide a robust algorithm for high Reynolds numbers. This choice is motivated by the fact that writing the solution in the \ac{TSE} form, or \ac{SPGD} form while considering $\T_\n(t) = t^\n$, renders the main problem into a cascade of linearized ones and skips the iterative and costly process in finding $\X_\n$. The \ac{NS} equations are given in their dimensionless form:
\begin{equation}
 \label{eq_NS_main}
 \begin{aligned}
 \bu_t + (\bu\cdot\nabla) \bu + \nabla \bp &= \frac{1}{\Rey} \Delta \bu ,\\
 \nabla \cdot \bu &= 0
 \end{aligned}
\end{equation}
where $\bu$ is the velocity vector field, $\bp$ is the scalar pressure, $\Rey$ is the dimensionless Reynolds number and $\nabla \cdot$ is the divergence operator. After injecting forms of \cref{u_p_TSE} in \cref{eq_NS_main} and equating terms multiplying $t^k$, we find the recurrence formula generating $(\uk_{k},\pk_{k-1})$ as follows:
\begin{equation}
 \label{eq_serie_NS}
 \forall k\geqslant 1
\left\lbrace
 \begin{aligned}
 \Fns(k) &= 0,\\
  \nabla\cdot \uk_k & = 0,
 \end{aligned}\right.
\end{equation}
where
\begin{equation}
 \label{Fns}
  \Fns(k) = k \uk_{k} + \nabla \pk_{k-1} +\sum\limits_{r=0}^{k-1} (\uk_{r}\cdot\nabla)\uk_{k-r-1} - \frac{1}{\Rey}\Delta \uk_{k-1},
\end{equation}
Then $(\uk_k,\pk_{k-1})$ are the solution of $\Fns(k)=0$ with the divergence free condition. When solving the above set of equations up to rank $k=\nn$ in \ac{FEM}, it was shown (see \cite{Deeb-NS-25}) that a stabilization, by adding an artificial diffusion term $\staba_{k-1} \Delta \uk_k$ is crucial and efficient to improve stability. Adding it while considering $\staba_k = k\alpha_0$ is equivalent to solving the \ac{NS} alpha model \cite{nsalpha-01,nsalpha-03}. For more details, see \cite{Deeb-NS-25}. The new recurrence formula of the \ac{STSE} generating $(\uk_{k},\pk_{k-1})$ is given below:
\begin{equation}
 \label{eq_serie_stab_NS}
 \forall k\geqslant 1
\left\lbrace
 \begin{aligned}
  \Fns(k) &=  \Gtse(k) ,\\
  \nabla\cdot \uk_k & = 0.
 \end{aligned}\right.
\end{equation}
We present in \cref{Alg_3} a pseudo algorithm that finds the $k^{th}$ rank velocity $\uk_k$ and $(k-1)^{th}$ rank pressure $\pk_{k-1}$ via the variational formulation. After building the space of velocity $V$ and pressure $Q$, we consider a Mixed formulation of the problem. Hereafter the variational formulation of system \eqref{eq_serie_stab_NS} after multiplying it by test functions $(v,q) \in W$:
\begin{equation}
\label{weak_form_NS_TSE}
\left\lbrace
\begin{array}{rcl}
 k\langle \uk, v\rangle  - \langle  \pk, \nabla\cdot v\rangle +  \staba_k \langle  \nabla \uk, \nabla v \rangle  &=& -\frac{1}{\Rey}\langle  \nabla \uk_{k-1}, \nabla v \rangle \\
 &&+ \sum\limits_{r=0}^{k-1} \langle (\uk_{r}\cdot\nabla)\uk_{k-r-1} ,v \rangle. \\
 \langle \nabla\cdot \uk,q\rangle &=& 0
\end{array}
\right.
\end{equation}
The solution of system \eqref{weak_form_NS_TSE} is the weak solution of $(\uk_k,\pk_{k-1})$.

\begin{algorithm}[!ht]
\caption{Compute $(\uk_\n,\pk_{\n-1})$ with optional stabilization.\label{Alg_3}}
\begin{algorithmic}[1]
\Require $\uk_0 = \bu(0,\bx)$; $\n$, integer; $W = V\times Q$ Space function
\Procedure{Compute\_n}{$\uk_0, \n$}
  \For{$k \gets 1$ \textbf{to} $\n$}
    \State $\uk,\pk\gets$ trial functions of $W$, $v,q \gets $ test functions of $W$
    \State $a_k \gets k \langle \uk,v \rangle - \langle  \pk, \nabla\cdot v\rangle + \staba_k\langle  \nabla \uk, \nabla v\rangle + \langle \nabla\cdot \uk,q\rangle$
    \State $L_k \gets -\frac{1}{\Rey}\langle  \nabla \uk_{k-1}, \nabla v \rangle$
    \For{$r \gets 0$ \textbf{to} $k-1$}
      \State $L_k\gets L_k +\langle (\uk_{r}\cdot\nabla)\uk_{k-r-1}  ,v\rangle$
    \EndFor
    \State $(\uk_k,\pk_{k-1}) \gets$ \textbf{solve} $a_k = L_k$ \textbf{with} Boundary conditions
  \EndFor
  \State \textbf{return} $(\uk_\n, \pk_{\n-1})$
\EndProcedure
\end{algorithmic}
\end{algorithm}

On the other hand, the \ac{SPGD} considers approximating the solution as presented in \cref{u_p_alt}, following the same procedure of finding  $\Xu_p$ and $\Xp_p$ with the orthogonality condition. We replace it with the approximations of the time and space derivative of $\bu$ with those of $\bp$ in \cref{eq_NS_main}  by summing up to rank $\n$ for the velocity $\spu$ and up to rank $\n-1$ for the pressure $\spp$,
multiplying by $t^\n$ and integrating over $[0,\rangeT]$, assembling terms of $\rangeT^{p},\, p=0,\ldots 2\n$, then dividing by $\frac{\rangeT^{2\n}}{2\n}$ and arranging terms, we got the equation governing $(\Xu_\n, \Xp_{\n-1})$ as follows:
\begin{equation}
 \label{eq_NS_alt1}
 \forall \n\geqslant 1
 \left\lbrace
 \begin{aligned}
\Fns(\n)&= \Gns(\n) + \sum\limits_{p=1}^{\n-1} \Fns(p) \,\wnp{p}{\n}\\
  \nabla\cdot \Xu_\n & = 0,
 \end{aligned}\right.
\end{equation}
where
Note that $\Fns(\n)$ is already defined in \cref{Fns} and:
\begin{equation}
\begin{aligned}
 \Gns(n) &= \stab_\n \Delta \Xu_\n  + \sum\limits_{p=0}^{\n} \left(\sum\limits_{r=0}^{\n-p} \left(\Xu_{p+r}\cdot \nabla\right)\Xu_{\n-r}  \right) \vnp{p}{\n}, \\
 \stab_\n  &=  \frac{2\n \rangeT}{(2\n+1) \Rey}, \quad
 \vnp{p}{\n} = - \frac{2\n\rangeT^{p+1}}{(2\n+p +1) }, \quad
 \wnp{p}{\n} = - \frac{2\n}{(\n+p) \rangeT^{\n-p}}.
\end{aligned}
\end{equation}
Solving this equation in \ac{FEM} framework is costly for assembling $\Fns(p)$ for $p=1,\ldots,\n-1$ and $\Gns(\n)$. Using the relation in \cref{formula_fn}, we can write the recurrence formula in \cref{eq_NS_alt1} as follows:
\begin{equation}
 \Fns(\n) = \sum\limits_{p=1}^{\n} \bnp{p}{\n}\, \, \Gns(p) = \sum\limits_{p=1}^{\n} \psi(p,\n)\, \, \Gns(p)
\end{equation}
The use of this formula is essential as it reduces the complexity of  \cref{eq_NS_alt1} and compacts the $\n^{th}$ rank of the momentum equation governing $(\Xu_\n,\,\Xp_{\n-1})$:
\begin{equation}
 \label{eq_NS_alt_final}
 \begin{aligned}
 \n \Xu_\n + \nabla \Xp_{\n-1} - \Gns(\n) &= - \sum\limits_{r=0}^{\n-1} \left(\Xu_{r}\cdot \nabla\right)\Xu_{\n-1-r}  + \frac{1}{\Rey} \Delta \Xu_{\n-1}+\sum\limits_{p=1}^{\n-1} \psi(p,\n) \, \Gns(p),
 \end{aligned}
\end{equation}
which is completed by the $\n^{th}$ rank of the divergence-free condition $\nabla\cdot \Xu_\n = 0$. \cref{Alg_4} presents the pseudo-algorithm of computing $(\Xu_\n,\Xp_{\n-1}) \in W$.
\begin{algorithm}[!ht]
\caption{Compute $\Xu_\n,\Xp_{\n-1}$.\label{Alg_4}}
\begin{algorithmic}[1]
\Require $\Xu_0 = \bu(0,\bx)$; $\n$, integer; $W = V\times Q$ Space function
\Procedure{Compute\_Xn}{$\Xu_0, \n$}
  \For{$p \gets 1$ \textbf{to} $\n$}
    \State $\Xu,\Xp\gets$ trial functions of $W$, $v,q \gets $ test functions of $W$
    \State $a_p \gets p \langle \Xu,v \rangle - \langle  \Xp, \nabla\cdot v\rangle - \langle\Gns(p),v\rangle   + \langle \nabla\cdot \Xu,q\rangle$
    \State $B_p \gets \text{Compute Path Sum}(p,\wnp{j}{p})$
    \State $L_p \gets -\frac{1}{\Rey}\langle  \nabla \Xu_{p-1}, \nabla v \rangle$
    \For{$r \gets 0$ \textbf{to} $p-1$}
      \State $L_p\gets L_p +\langle (\Xu_{r}\cdot\nabla)\Xu_{p-r-1}  ,v\rangle$
    \EndFor
    \For{$r \gets 1$ \textbf{to} $p-1$}
      \State $L_p\gets L_p +B_p[r]\, \langle \Gns(r)  ,v\rangle$
    \EndFor

    \State $(\Xu_p,\Xp_{p-1}) \gets$ \textbf{solve} $a_p = L_p$ \textbf{with} Boundary conditions
  \EndFor
  \State \textbf{return} $(\Xu_\n, \Xp_{\n-1})$
\EndProcedure
\end{algorithmic}
\end{algorithm}

We can summarize the connection between \ac{TSE}, \ac{STSE} and \ac{SPGD} as follows:
\begin{equation}
 \Fns(\n)  = \left\lbrace
 \begin{array}{l|r}
  0, & \ac{TSE} \\
  \Gtse(\n),   & \ac{STSE} \\
  \sum\limits_{p=1}^{\n} \bnp{p}{\n}\, \, \Gns(p),  & \ac{SPGD}
 \end{array}
 \right.
\end{equation}
This connection is essential to understand the stabilization technique, allowing us to choose $\stab_k$ as closed forms of the coefficients of stabilization $\staba_k$ in the case of \ac{STSE}.
Next, we will use both methods, \ac{STSE} and \ac{SPGD}, to simulate the two-dimensional wake behind a bluff body.

\section{Simulating the Wake of a 2D Bluff Body}
\label{sec5}
To assess the performance and practical relevance of the proposed methods, we apply both the \ac{STSE} and \ac{SPGD} formulations to the simulation of the unsteady flow around a two-dimensional bluff body governed by the incompressible Navier–Stokes equations. This test case is a classical benchmark in fluid dynamics due to the rich variety of flow structures it produces, especially in the wake region.

\subsection{Geometric features}
\begin{figure}[h!]
 \centering
 \includegraphics[scale=0.35]{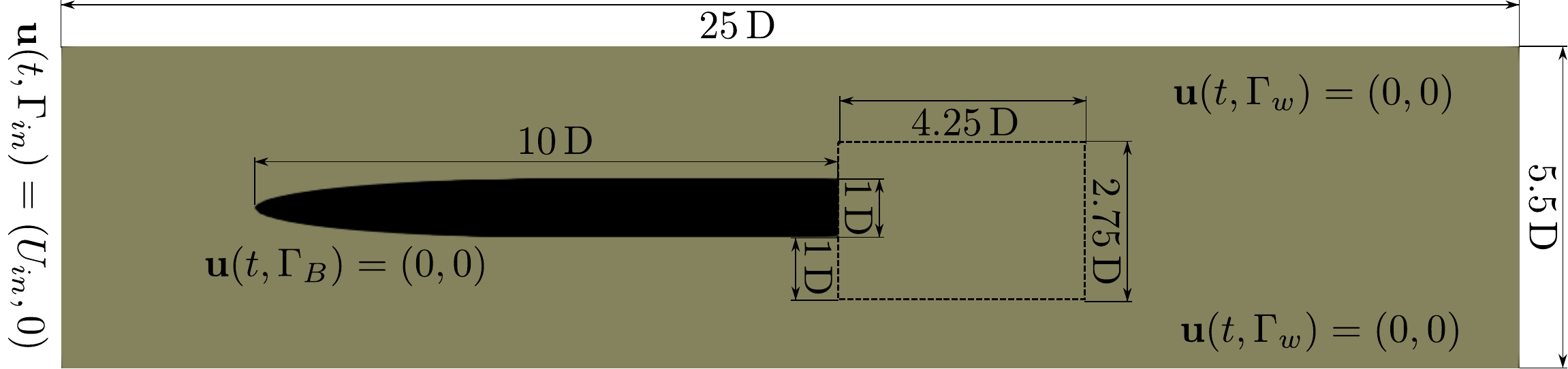}
 \caption{Illustration of the bluff body geometry in the flow.\label{Fig_geo_bluff}}
\end{figure}

We consider the value $\D$ as the length characteristic. The bluff body, denoted by $B$, has an elongated D-shape bluff body with a base equal to $\D$ and $10\,\D$ of length. It is a combination of two geometric shapes: $B= S_1\cup S_2$. The first part $S_1$, facing the flow, is formed of a semi ellipse centred at $(x_c = \frac{25}{3}\D,\,y_c =\frac{11}{4}\,\D )$, with central axis $5\,\D$ and minor axis $\frac{1}{2}\D$, while the second part $S_2$ is a rectangle with a length of $5\,\D$ and width $\D$. The following sets represent both parts:
\begin{equation}
\begin{aligned}
 S_1 &= \left\lbrace(x,y) \left\vert\frac{ \left(x-x_c\right)^2}{(5\D)^2} + \frac{\left(y-y_c\right)^2}{(\frac{1}{2}\D)^2} \leqslant 1,\right. \,\,  x\leqslant x_c \right\rbrace, \\
 S_2 &= \left\lbrace (x,y) \left\vert x_c\leqslant x\leqslant x_c+5\,\D,\,\&\, y_c-\D \leqslant y \leqslant y_c+\frac{1}{2}\D \right. \right\rbrace.
 \end{aligned}
\end{equation}
The bluff body is placed in a rectangular domain $\Omega$ having a width equal to $5.5\,\D$ and a length of $25\,\D$. The geometric features and boundary conditions are described in \cref{Fig_geo_bluff}. This geometry is chosen to prepare for future work to compare the three-dimensional numerical simulation obtained with the proposed method and experimental data acquired with different Reynolds numbers in the wake region.

The flow is initiated from the left side of the rectangle, \emph{i.e.} $\Gamma_{in} = \{(x,y)\in\,\mathds{R}^2, \rvert x=0, 0\leqslant y\leqslant 5.5\, \D\}$, with a velocity inflow $(U_{in},0)^\top$. The dashed rectangle downstream of the bluff body represents the wake region. The boundary conditions on the upper and lower walls, \emph{i.e.} $\Gamma_w = \{(x,y) \rvert 0\leqslant x\leqslant 25\,\D, y=0,\,5.5\,\D \}$, are no-slip boundary conditions, \emph{i.e.} $\bu(t,\Gamma_w) = (0,0)$, same as the conditions on the boundary $\Gamma_B$ of the bluff body. The outflow (right side of $\Omega$) is a free-stream flow condition. The flow is governed by the dimensionless \ac{NS} equation with $\Rey =  \D\,\rho\, U_{in} / \mu$, where $\nu =\mu/\rho$ is the kinematic viscosity of the air. The dimensionless system is presented below.
\begin{equation}
\label{NS_BB_Eq_adim}
 \begin{aligned}
  \bu_t + (\bu\cdot\nabla)\bu + \nabla \bp &= \frac{1}{\Rey} \Delta \bu, & (t,\bx) \in [0,\mT]\times\Omega\backslash B,\\
  \nabla\cdot \bu & = 0, & (t,\bx) \in [0,\mT]\times\Omega\backslash B,\\
 \bu(t,\bx) &= (1,0)^\top, & (t,\bx) \in [0,\mT]\times\Gamma_{in},\\
 \bu(t,\bx) &= (0,0)^\top, & (t,\bx) \in [0,\mT]\times(\Gamma_{w}\cup \Gamma_B).
 \end{aligned}
\end{equation}
The above dimensionless system is obtained after considering the following variable change, where the superscript $\star$ corresponds to the dimensional variables:
\begin{align*}
 \bxs &= \D\,\bx,& \bus &= U_{in}\bu, & \ts& = \frac{\D}{U_{in}} t,& \bps &= \rho U_{in}^2 \bp.
\end{align*}

\subsection{Numerical features}
After building the geometry, the mesh is built using \textit{Gmsh} (4.8.3) \cite{gmsh-09} for a mesh precision parameter $N_x$, then loaded in \textit{Python} to run the simulation using the \ac{FEM} solver in \textit{FEniCS}  (2019.X) \cite{LoggWells2010,fenics2012,AlnaesEtal2014,AlnaesEtal2015}. The space of function is a Mixed space $\M = \V\times \Q$ of the velocity vector space function $\V$ and the pressure scalar space function $\Q$. The velocity space $\V$ is the vector space of dimension two, where each dimension is the set of functions that are second-order polynomials when projected on every element in the mesh. In contrast, the first-order polynomials generate $\Q$. The initial condition is considered to be the stationary solution of the Stokes problem with $\Rey= 1000$. The simulation is performed over the interval $[0,\mT]$ where $\mT = 2.5\frac{ U_{in}}{\D}$ representing the dimensionless time.

\subsection{Numerical Setup}
\begin{figure}[h!]
 \centering
 \includegraphics[scale=0.22]{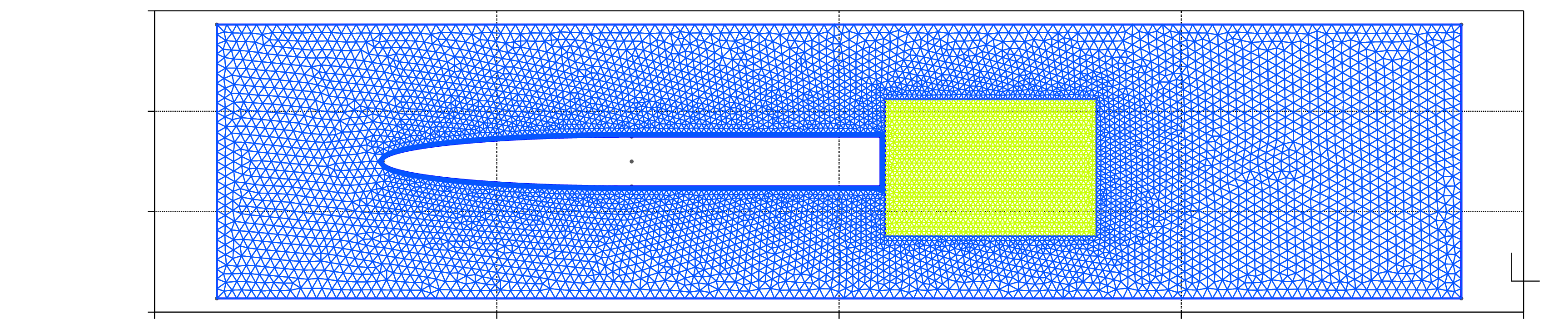}
 \caption{The mesh used in the simulation. \label{mesh_body2_NAS2}}
\end{figure}

This section presents the numerical results of the simulation run by the \ac{STSE} and \ac{SPGD} methods. The mesh is built using \text{Gmsh} software. A boundary layer is built around the body using the Field \textrm{BoundaryLayer} with a ratio parameters $growth_{rate} = 1.2$ and $hwall_n$ parameters is equal to  $= 0.02\D \cdot 1.2^{5}$ where $thickness = 0.02 \D$, $growth_{rate} = ratio$ and $num_{layers} = 6$. The wake zone (color green in \cref{mesh_body2_NAS2}) is meshed using a Transfinite function.
After meshing, we obtain a mesh with 9020 nodes and 18194 elements (see \cref{mesh_body2_NAS2}). The minimum cell diameter equals tp $0.004916$ and the maximum is $0.025351$.

\begin{figure}[h!]
 \centering
 \includegraphics[scale=0.34]{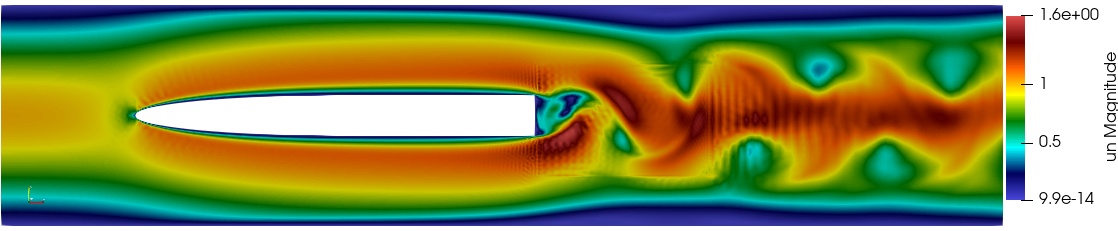}
 \includegraphics[scale=0.34]{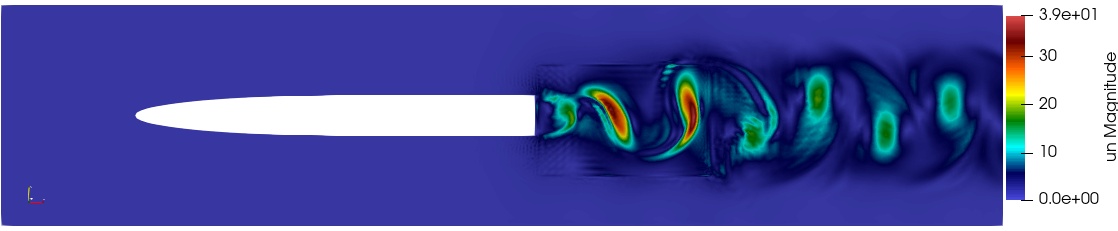}
 \includegraphics[scale=0.34]{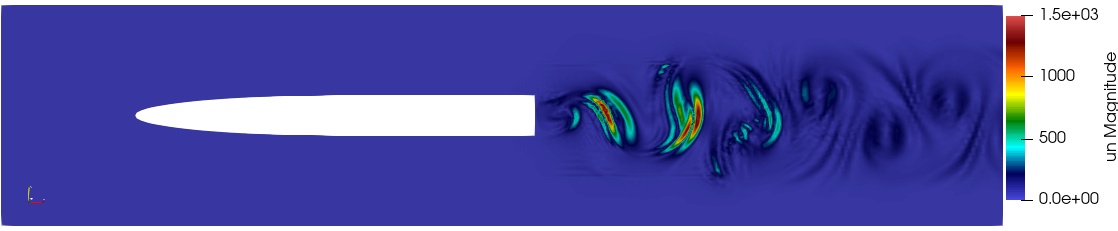}
 \includegraphics[scale=0.34]{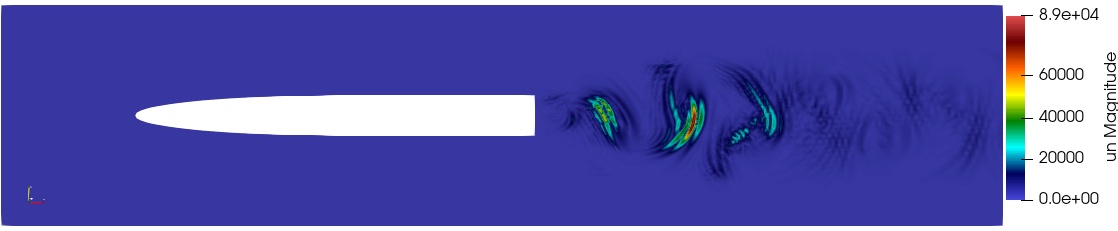}
 \caption{Plot of the first four (from up to down) space modes $\uk_0,\uk_1,\uk_2,\uk_3$ of the numerical simulation with \ac{STSE} method at $t=4$. \label{fig_mode_un0123}}
\end{figure}

\begin{figure}[h!]
 \centering
  \includegraphics[scale=0.34]{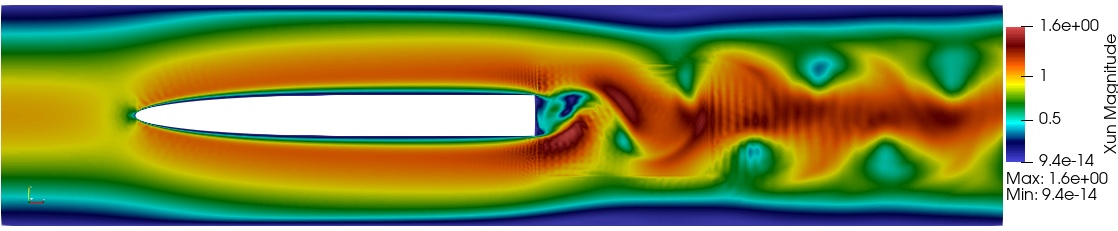}
  \includegraphics[scale=0.34]{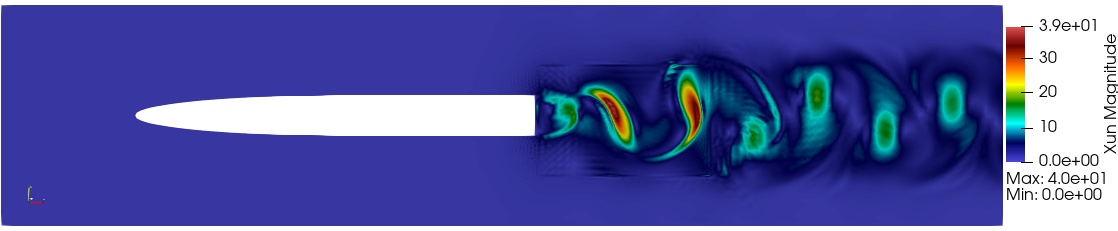}
 \includegraphics[scale=0.34]{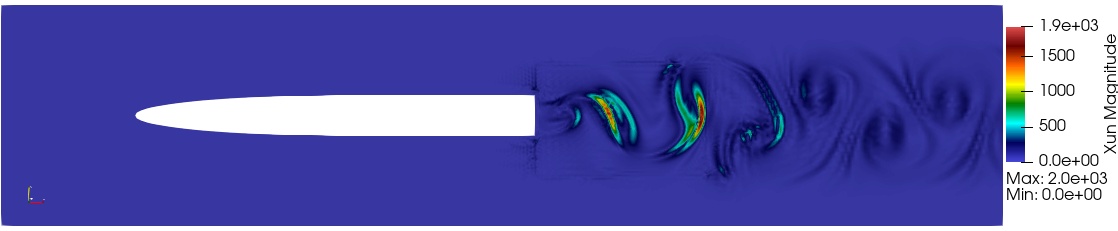}
 \includegraphics[scale=0.34]{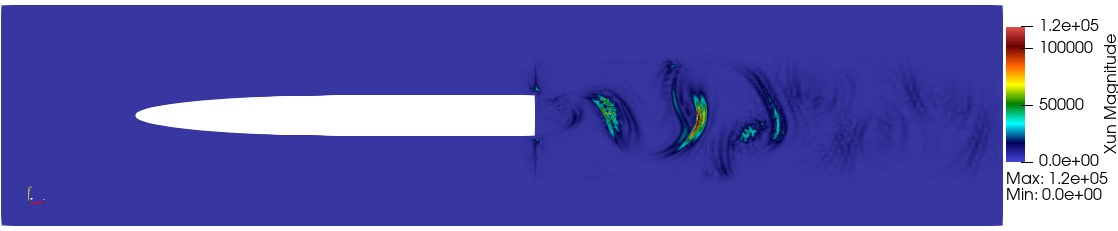}
 \caption{Plot of the first four (from up to down) space modes $\mX_0,\mX_1,\mX_2,\mX_3$ of the numerical simulation with \ac{SPGD} method at $t=4$. \label{fig_mode_xun0123}}
\end{figure}

The time step is taken to be equal to $\rangeT = 4\cdot10^{-3}$. The initial condition is considered to be the solution of the Stokes problem. The stabilization coefficients $\staba_{k}$ of the \ac{STSE} method are equal to $\stab_k$. After reaching an oscillating regime as shown in the first image in \cref{fig_mode_un0123}, we plot the first four modes of the \ac{STSE} method when $t=4$. We can see that the series diverges as the magnitude of the space modes increases with $k$. These modes are also plotted in \cref{fig_mode_xun0123} for the \ac{SPGD} solver.

\begin{figure}[h!]
 \centering
 \includegraphics[scale=0.4]{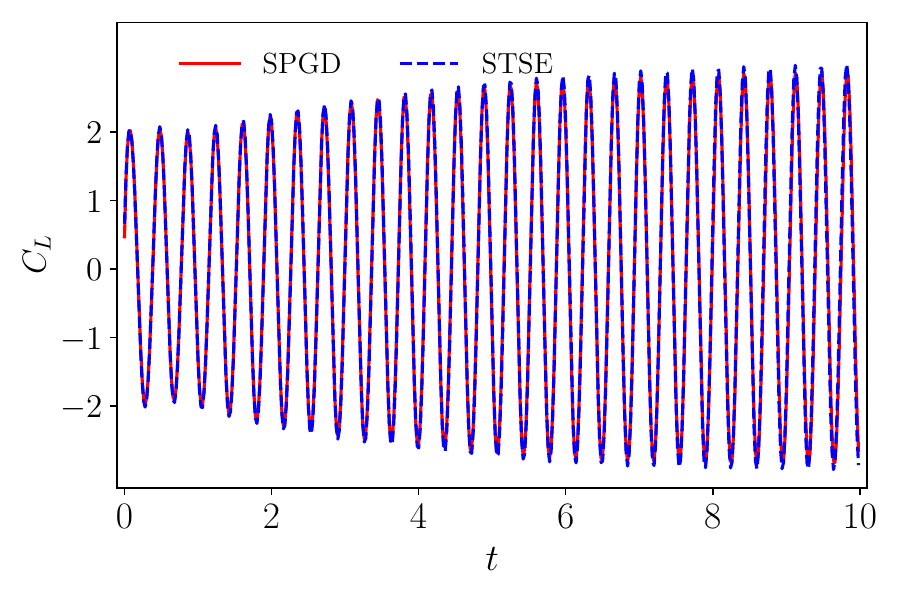}
 \includegraphics[scale=0.4]{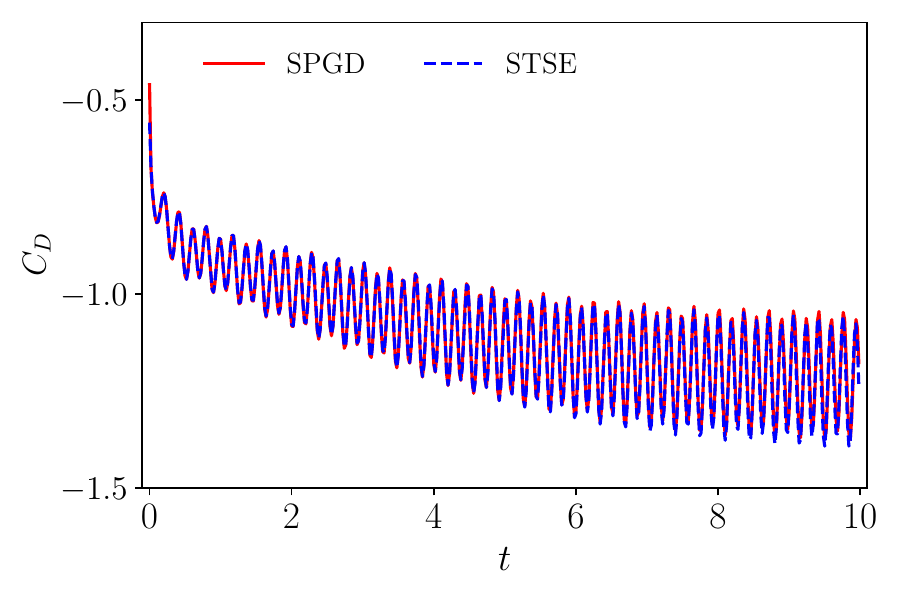}
 \caption{Lift and Drag coefficients using \ac{SPGD} and \ac{STSE}. \label{fig_Lift_drag_coef}}
\end{figure}

We compute the drag and lift coefficients based on the fluid stress tensor evaluated on the body surface to quantify the hydrodynamic forces acting on the bluff body. The total stress tensor $\boldsymbol{\sigma}$ is defined as:

\begin{equation}
\boldsymbol{\sigma} = -\bp \I + \mu \left( \nabla \bu + (\nabla \bu)^\top \right),
\end{equation}
where $\bp$ is the pressure, $\mu$ is the dynamic viscosity, and $\bu$ is the velocity field. The traction vector (force per unit area) on the surface is then given by:
\begin{equation}
\boldsymbol{f} = \boldsymbol{\sigma} \cdot \boldsymbol{n},
\end{equation}
where $\boldsymbol{n}$ is the outward unit normal to the body surface. The drag and lift forces are computed as:
\begin{align}
F_x &= \int_{\Gamma_{B}} (\boldsymbol{\sigma} \cdot \boldsymbol{n}) \cdot \hat{\boldsymbol{x}} \, ds, &
F_y &= \int_{\Gamma_{B}} (\boldsymbol{\sigma} \cdot \boldsymbol{n}) \cdot \hat{\boldsymbol{y}} \, ds,
\end{align}
where $\hat{\boldsymbol{x}}$ and $\hat{\boldsymbol{y}}$ are the unit vectors in the stream-wise (drag) and transverse (lift) directions, respectively. The dimensionless drag and lift coefficients are then computed as:
\begin{equation}
C_D = \frac{F_x}{\frac{1}{2} \rho U_{in}^2 A}, \quad
C_L = \frac{F_y}{\frac{1}{2} \rho U_{in}^2 A},
\end{equation}
where $\rho$ is the fluid density, $U_{in}$ is the reference inlet velocity, and $A$ is the frontal area of the bluff body. In our two-dimensional setup, the frontal area is defined as $A = 2H / L_{\text{char}}$, where $H = 2\,\D = 50 \times 10^{-3}$ m is the body height. We plot these coefficients in \cref{fig_Lift_drag_coef} obtained by both methods: \ac{SPGD} and \ac{STSE}. We can see that both have the same pattern of oscillating.
We also simulate with a smaller time step $\rangeT =  2\cdot10^{-3}$ to check the quality of the simulation. We conclude that both methods have shown what was shown above in the last section: the performance of these methods maintains their stability for a range of time steps $\rangeT$.

\begin{figure}[h!]
 \centering
 \includegraphics[scale=0.4]{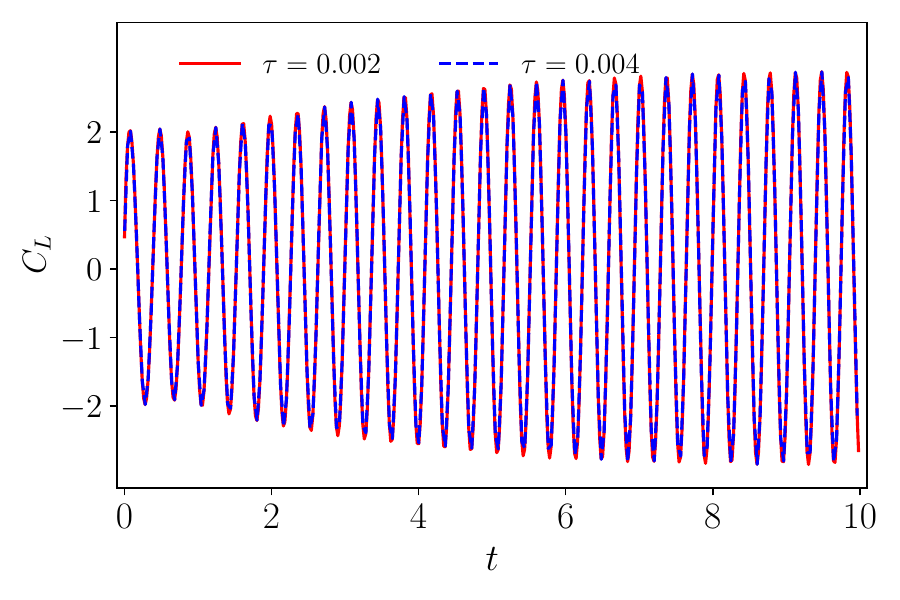}
 \includegraphics[scale=0.4]{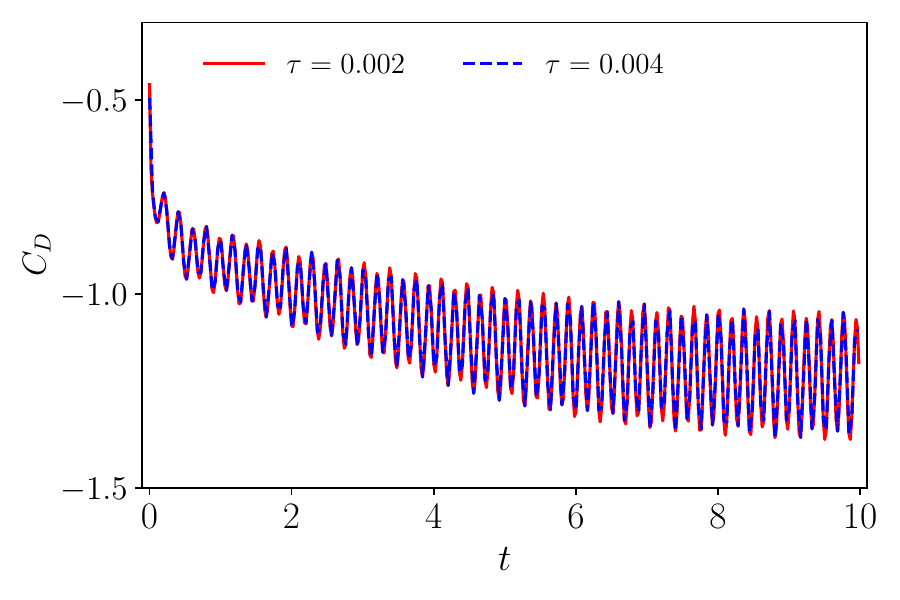}
 \caption{Lift and Drag coefficients with two values of $\rangeT=0.002,0.004$ using \ac{SPGD}. \label{fig_Lift_drag_coef_pgd}}
\end{figure}

\section{Conclusion}
\label{sec6}
In this paper, we presented a graph-based framework that connects methods based on the \acf{TSE}, including its stabilized variant \ac{STSE}, to the \acf{PGD} method. Using graph theory, we reduce the complexity of computing space modes in the PGD formulation. In particular, when we represent time modes in PGD using a \ac{TSE} structure, the resulting two-level Volterra-type recurrence relation, governing time modes coefficients, can be simplified through the concept of simple paths in a directed graph, especially under the assumption that the initial space tensor coincides with the initial condition. We presented a theorem allowing us to construct the \ac{TSE} space modes via the \ac{PGD} space modes without the Picard iterative process.

On the other hand, we showed that the computation of space tensors in \ac{PGD} reveals a natural stabilization mechanism through artificial diffusion. This observation supports and strengthens our earlier work, where we proposed a stabilization technique for the TSE method by introducing artificial diffusion.

We also proposed a method that bridges \ac{STSE} and \ac{PGD}, termed the \acf{SPGD}, in which we choose the time tensors as monomials $t^n$. By leveraging graph theory again, we derived a simplified one-level Volterra-type recurrence formula for the space tensors. The resulting formulation yields stabilization coefficients naturally, as functions of the time step $\tau$, the tensor rank $\n$, and the governing \ac{PDE}.

We assessed the numerical performance of the \ac{STSE} methods (with two different sets of stabilization coefficients) and \ac{SPGD} through simulations of the diffusion equation under varying mesh sizes $h$, time steps $\rangeT$, finite element orders $s$, and diffusivity values $\nu$. The results show that the \ac{STSE} method, using stabilization coefficients derived from \ac{SPGD}, exhibits greater robustness than the version using coefficients $\staba_0$, previously obtained via offline minimization of the mass matrix condition number \cite{deeb:stab-serie}.

We extend the graph-based connection framework to inviscid flow, aiming to see how crucial it is to add an artificial diffusion term in the right-hand side of the recurrence formula for stabilizing the simulation for large-time simulation, allowing time steps 100 times larger.

Finally, we compare the \ac{STSE} and \ac{SPGD} to simulate the dimensionless incompressible Navier–Stokes equations. After deriving the space tensor recurrence relations for both \ac{STSE} and \ac{SPGD}, showing the connection, we applied them to simulate the wake behind a bluff body in the framework of \ac{FEM} using FEniCS. We computed and compared the drag and lift coefficients obtained from both methods under various simulation settings. The results demonstrated stability and physical consistency for both approaches.

\section*{Acknowledgement}
This work has been supported by the Khalifa University of Science and Technology under Award No. RIG-2023-024. This work is based upon work supported by the Khalifa University of Science and Technology under Award No. FSU-2023-014.

\appendix
\section{Algorithm of computing $\psi(p,n)$ }
\label{append_alg_bnp}
\begin{algorithm}[!ht]
	\caption{Compute coefficients $\bnp{p}{\n}\coloneqq \psi(p,n)$ via signed path sums. \label{Alg_bnp}}
	\begin{algorithmic}[1]
		\State \textbf{Input:} $\n$, $p$, integer; $\wnp{p}{\n}$, array of weights
		\State \textbf{Output:} $B_\n$, array of $\bnp{p}{\n}$
		\Procedure{Compute Path Sum}{$\n,\wnp{p}{\n}$}
		\State Initialize directed graph $DiGr$
		\For{$i \gets 1$ \textbf{to} $\n-1$}
		\For{$j \gets i+1$ \textbf{to} $\n$}
		\State Add edge $(i \to j)$ to $DiGr$ with weight $\wnp{i}{j}$
		\EndFor
		\EndFor
		\State $B_\n$ zero array of length $\n$
		\For{$p \gets 1$ \textbf{to} $\n-1$}
		\State total $\gets 0$
		\ForAll{simple paths $P=(p \to \cdots \to \n)$ in $DiGr$}
		\State product $\gets \prod\limits_{(i \to j)\in P} \wnp{i}{j}$
		\State total $\gets$ total $+$ product
		\EndFor
		\State $\bnp {p}{\n}\gets$ total
		\State $B_\n[p] \gets \bnp{p}{\n}$
		\EndFor
		\State \textbf{return} $B_\n$
		\EndProcedure
	\end{algorithmic}
\end{algorithm}

\begin{acronym}[TDMA]
\acro{DNS}{Direct numerical simulations}
\acro{DMP}{Discrete Maximum Principle}
\acro{FE}{Finite Element}
\acro{NS}{Navier-Stokes}
\acro{RT}{Raviart-Thomas}
\acro{FSI}{Fluid structure Interaction}
\acro{TSE}{Time Series Expansion}
\acro{STSE}{Stabilized-\ac{TSE}}
\acro{FEM}{Finite Element Method}
\acro{FDM}{Finite Difference Method}
\acro{PGD}{Proper Generalized Decomposition}
\acro{SPGD}{Simplified \ac{PGD}}
\acro{POD}{Proper Orthogonal Decomposition}
\acro{PDE}{Partial Differential Equation}
\acro{ODE}{Ordinary Differential Equation}
\acro{IVP}{Initial Value Problem}
\acro{DSR}{Divergent Series Resummation}
\acro{PIV}{Particle Image Velocimetry}
\acro{LES}{Large Eddy Simulation}
\acro{SUPG}{Streamline Upwind/Petrov-Galerkin}
\acro{PSPG}{Pressure Stabilized/Petrov-Galerkin}
\acro{BDF}{Backward Difference Formulas}
\end{acronym}

\bibliographystyle{elsarticle-num-names}
\bibliography{biblio}

\end{document}